\documentclass[twocolumn]{emulateapj}
\usepackage{amstext}
\usepackage{apjfonts}
\usepackage{amsmath}
\usepackage{amssymb}
\usepackage{enumerate}
\usepackage{graphicx}
\usepackage{xcolor}
\usepackage{xspace}
\usepackage[colorlinks=true,linkcolor=blue,citecolor=blue]{hyperref}

\definecolor{darkgreen}{rgb}{0,0.5,0}
\definecolor{darkyellow}{rgb}{0.75,0.75,0}

\defcitealias{Gezari:2012fk}{G12}
\defcitealias{Guillochon:2013jj}{GR13}
\newcommand{\dmde}{dM/dE}
\newcommand{\md}{\dot{M}}
\newcommand{\medd}{\dot{M}_{\rm Edd}}
\newcommand{\mdpeak}{\dot{M}_{\rm peak}}

\newcommand{\msun}{M_{\odot}}

\newcommand{\ha}{{\rm H}$\alpha$\xspace}
\newcommand{\hb}{{\rm H}$\beta$\xspace}
\newcommand{\hg}{{\rm H}$\gamma$\xspace}
\newcommand{\Heii}{\ion{He}{2}\xspace}
\newcommand{\heii}{\ion{He}{2} $\lambda$4686\xspace}
\newcommand{\hei}{\ion{He}{1} $\lambda$5876\xspace}
\newcommand{\psone}{PS1-10jh\xspace}

\newcommand{\tdefit}{{\tt TDEFit}\xspace}
\newcommand{\FLASH}{{\tt FLASH}\xspace}

\newcommand{\Sim}{\mathord{\sim}}
\newcommand{\V}{{\cal V}}
\newcommand{\Log}{{\rm Log}\xspace}
\newcommand{\rsep}{\leq x \leq}

\begin{document}

\shortauthors{Guillochon, Manukian, Ramirez-Ruiz}

\title{PS1-10jh: The Disruption of a Main-Sequence Star of Near-Solar Composition}

\author{James Guillochon, Haik Manukian, and Enrico Ramirez-Ruiz \altaffilmark{1}}
\altaffiltext{1}{Department of Astronomy and
  Astrophysics, University of California, Santa Cruz, CA
  95064}
   
\begin{abstract} 
When a star comes within a critical distance to a supermassive black hole (SMBH), immense tidal forces disrupt the star, resulting in a stream of debris that falls back onto the SMBH and powers a luminous flare. In this paper, we perform hydrodynamical simulations of the disruption of a main-sequence star by a SMBH to characterize the evolution of the debris stream after a tidal disruption. We demonstrate that this debris stream is confined by self-gravity in the two directions perpendicular to the original direction of the star's travel, and as a consequence has a negligible surface area and makes almost no contribution to either the continuum or line emission. We therefore propose that any observed emission lines are not the result of photoionization in this unbound debris, but are produced in the region above and below the forming elliptical accretion disk, analogous to the broad-line region (BLR) in steadily-accreting active galactic nuclei. As each line within a BLR is observationally linked to a particular location in the accretion disk, we suggest that the absence of a line indicates that the accretion disk does not yet extend to the distance required to produce that line. This model can be used to understand the spectral properties of the tidal disruption event (TDE) \psone, for which \ion{He}{2} lines are observed, but the Balmer series and \ion{He}{1} are not. Using a maximum likelihood analysis, we show that the disruption of a main-sequence star of near-solar composition can reproduce this event.
\end{abstract}


\section{Introduction}\label{sec:intro}

The tidal disruption of a star by a supermassive black hole (SMBH) splits the star into either two or three ballistically distinct masses. In the event of a full disruption, the star is split into two pieces of nearly-equal mass. One half of the star becomes bound to the black hole after the encounter, and continues along elliptical trajectories with pericenter distances equal to the star's original pericenter distance. The other half of the star gains orbital energy in the encounter, and is placed on hyperbolic trajectories. For a partial disruption, a third mass in the form of a surviving stellar core emerges from the encounter, with the absolute value of its orbital energy comparable to its own binding energy \citep{Faber:2005be,Guillochon:2011be,MacLeod:2012cd,Liu:2013er,Manukian:2013ce}.

Determining the fates of these pieces of the star are critical in determining the appearance of the flare that results from the immense gravitational energy that will be released by the accretion disk that eventually forms. Previously, it has been assumed that the unbound material, which was thought to be a wide ``fan,'' was the primary contributor to the broad emission lines that are produced as the result of a tidal disruption \citep{Strubbe:2009ek,Kasen:2010ci,Clausen:2011fa}.

For the tidal disruption event (TDE) \psone (\citealt{Gezari:2012fk}, hereafter \citetalias{Gezari:2012fk}), it was assumed that hydrogen, which is ejected to large distances within the wide debris fan generated by the disruption, can recombine more quickly than the rate at which it is ionized by the central source. This would ensure that the vast majority of the hydrogen is neutral, and thus any ionizing radiation incident upon the fan would produce an emission feature. The absence of any hydrogen emission features was used to derive an upper limit on the amount of hydrogen present, implying that helium is five times more common than hydrogen by mass with the disrupted star.

In this paper, we present three-dimensional hydrodynamical simulations that show that the assumption that this debris fan intercepts a significant fraction of the light is incorrect. As noted by \cite{Kochanek:1994bn}, the width of the stream of unbound material is still controlled by the stream's self-gravity in the transverse direction, restricting its width to only be a fraction of the star's original pericenter distance. Through numerical simulations of fully-disruptive encounters with mass ratios $q \equiv M_{\rm h}/M_{\odot} = 10^{3}$ and $10^{6}$, we verify that the transverse containment of the stream's width does indeed occur. As a result, the stream only grows in the radial direction, and thus the total volume and surface area increase only slightly more steeply than $v_{\rm p}$. Therefore, the emitting volume of this structure is not significant enough to produce bright hydrogen emission lines, even for the disruption of a main-sequence (MS) star composed largely of hydrogen.

But while we find that the area of the unbound debris has been vastly overestimated, we also find that the area occupied by the accretion disk formed from the bound material has been vastly underestimated. Our numerical simulations confirm the prediction that material that returns to pericenter is ballistically launched to very large distances from the black hole, hundreds of times $r_{\rm p}$. Additionally, we find that significant dissipation occurs when this material returns to pericenter. As the debris stream quickly virializes at pericenter and the density of the material is significantly reduced as compared to the star's original density, self-gravity is suppressed even in the transverse direction. As a result, a fan structure {\it is} formed once material returns pericenter. But as this material belongs to the fraction of the original star that is strongly bound to the black hole, the radial extent of this material grows at a rate that is significantly smaller than the unbound fraction.

As the region in which \ha is produced in steady AGN is on the order of a few light days to a few light weeks from the black hole for $L_{5100} \sim 10^{45}$ ergs s$^{-1}$ \citep{Peterson:2006hr}, we show that it is unlikely that the debris ejected by the disruption has traveled the distance necessary to produce an \ha line for \psone. Through comparison with the processes responsible for producing the broad line regions (BLRs) of steadily-accreting AGN, we predict that the helium lines that are observed in \psone are produced much closer to the black hole \citep{Korista:1995eh,Bentz:2010hk}, and the debris has sufficient time to reach this distance by the time the first spectrum was observed. Motivated by the results of our hydrodynamical simulations, we model the accretion disk structure and use a Markov-Chain Monte Carlo (MCMC) procedure to determine the combinations of parameters with the highest-likelihood, and we find that the highest-likelihood models that do not include priors on the input parameters involve the disruption of a main-sequence star with mass $4 M_{\odot}$ by a $M_{\rm h} = 2 \times 10^{7} M_{\odot}$ black hole.

In Section \ref{sec:method} we describe our method for running hydrodynamical simulations to characterize the behavior of the debris stream after a disruption, and describe the maximum likelihood analysis (MLA) we employed to estimate the parameters of \psone. In Section \ref{sec:debris} we present a physical interpretation of the results of our hydrodynamical simulations. Bearing these results in mind, we  develop our generalized model for the time-dependent, broadband light that would accompany the disruption of a star in Section \ref{sec:model}. We then apply this model specifically to \psone in Section \ref{sec:ps1-10jh}. Finally, we review additional evidence as to why the disruption of a helium-rich star is unlikely to have produced \psone in the first place, and look towards the future when TDEs will be regularly observed.

\section{Method}\label{sec:method}

\subsection{Hydrodynamical Simulations}\label{subsec:hydrosims}
The black hole at the center of our own galaxy is estimated to be $\simeq 4 \times 10^{6} M_{\odot}$ \citep{Ghez:2008hf}, and is one of the smallest known massive black holes \citep{Schulze:2010jl}. As the mass of an average main-sequence star is $\sim 0.1 M_{\odot}$ \citep{Kroupa:1993tm}, the majority of stellar tidal disruptions will have $q \gtrsim 10^{6}$.  For such disruptions, the timescale of return of the most bound debris is on the order of days to weeks \citep{Rees:1988ei}, with the peak fallback rate occurring approximately one month after the time of the disruption (\citealt{Evans:1989jua}; \citealt{Lodato:2009iba}; \citetalias{Guillochon:2013jj}).

For hydrodynamical simulations of TDEs, the main limiting factor is the sound-crossing time of the star, which for a solar mass star is approximately one hour. Given an initial stellar model that occupies $100^{3}$ grid cells, each hydrodynamical time-step translates to only one minute of physical time. Thus, the simulation of the tidal disruption of a solar mass star by a $10^{6} M_{\odot}$ black hole that includes the time at which the fallback rate is at a maximum requires $\sim 10^{5}$ time-steps. Additionally, the debris stream resulting from the disruption must be fully resolved in both length and in width. As the stream is self-gravitating (as described in Section \ref{subsec:stream}), it mains a very narrow profile, with the aspect ratio of the stream when the first material returns to pericenter being $q^{1/3} (t_{\rm peak}/t_{\rm p})^{1/2} \sim 10^{3}$, where $t_{\rm peak}$ is the time where $\dot{M}$ reaches a maximum, and $t_{\rm p}$ is the pericenter crossing time. If the number of grid cells across the stream is forced to be at least 20, which is necessary to satisfy the \citet{Truelove:1997bj} criteria, then $10^{6}$ grid cells would be required to be evolved for $10^{5}$ time-steps.

This means that a complete simulation of the full problem within a single simulation is very computationally expensive. Instead, we run two separate simulations that are each well-equipped to describe the behavior of the debris stream at two different epochs. To determine the fate of the debris liberated from a star during a tidal disruption, we used two similar simulation setups, differing only in the mass ratio $q$. The first simulation sets $q = 10^{6}$ and solely focuses on the evolution of the debris stream as it expands away from pericenter after the star's initial encounter with the black hole. Because of the computational expense, the return of the debris to pericenter is not followed in this simulation. The second simulation sets $q = 10^{3}$, and follows the return of the debris to pericenter well beyond the peak in the accretion rate \smash{$\dot{M}_{\rm peak}$}. In these encounters, the peak accretion rate is realized only one day after pericenter, and we follow the evolution of the returning debris for a total of $5 \times 10^{5}$ seconds (about one week). Our hydrodynamical simulations were performed in a module written for the \FLASH adaptive mesh refinement code, the details of which can be found in \citet{Guillochon:2009di,Guillochon:2011be} and \citetalias{Guillochon:2013jj}.

The initial conditions of the simulation are similar to those presented in \citetalias{Guillochon:2013jj}, with the polytropic $\Gamma$ that describes the star's structure being set to 5/3, and the impact parameter $\beta \equiv r_{\rm p} / r_{\rm t}$ being set to 2. The star is placed on a parabolic trajectory at an initial position that is several times further than $r_{\rm t}$, and is initially resolved by 50 grid cells across its diameter. As realistic equations of state are only sensical at the full-scale of the problem, the hydrodynamics of the gas are treated using a simple adiabatic polytrope $P \propto \rho^{\gamma}$, where $\gamma$ is the adiabatic index. As we do not include any explicit viscosity terms, entropy generation only occurs through dissipation via shocks, the effects of which are captured by simultaneous evolution of the internal energy $\epsilon$. The code utilizes the adaptive-mesh functionality of the \FLASH software in different ways for the two simulations. In the $q = 10^{3}$ simulation, regions which are less than $10^{-1}$ times as dense as the current peak density are derefined, but maximum refinement is maintained within $4 r_{\rm p}$ at all times. For the $q = 10^{6}$ simulation, each refinement level is assigned to a single decade in $\rho$, using the star's original central density $\rho_{\rm c}$ as a baseline, with the exception of the first refinement threshold which is set to $\rho = 5 \times 10^{-3} \rho_{\rm c}$.

As both the timescales and the length scales of a $q = 10^{3}$ disruption are different from those of the more typical $q = 10^{6}$ disruption, care must be taken when interpreting the results from these simulations and attempting to scale them up to what would be realized for larger mass ratios. As we will describe in Section \ref{subsec:dissipative}, the dissipation processes that are observed in the scaled-down simulation are analogous to other dissipation mechanisms that operate for larger values of $q$, given the proper scaling.

\subsection{Fitting TDE Observations}

For fitting our models for tidal disruptions to observed events, we have developed the code \tdefit, which performs a MLA using an affine-invariant MCMC \citep{Goodman:2010et,ForemanMackey:2013io}, in which parameter combinations are assigned to individual ``walkers'' who then exchange positions according to their relative scores. The code is written in Fortran and utilizes the parallel variant of the algorithm presented in \citeauthor{ForemanMackey:2013io}. We have designed the software to be flexible in the model parameters it accepts as inputs, any free parameter (either discrete or continuous) can be included in the parameter space exploration by simply listing it and its range of acceptable values within a parameter file. In the same way, both trivial and non-trivial priors can be specified at runtime for single or combinations of input parameters.

As the solutions can often be multi-modal, with small regions of acceptable parameter space separated by large voids of poor parameter space, it can sometimes be difficult to find the deepest global minimum using the vanilla affine-invariant algorithm. To address this issue, we modify the algorithm by performing simulated annealing \citep[SA,][]{Press:1986vx} on a fraction $F$ of the walkers every $N$ timesteps during a ``bake-in'' period, where both $F$ and $N$ are adjustable. Each walker that is selected to anneal is used to seed an amoeba whose points are randomly drawn a small distance away from the original walker, these walkers then run through a full SA cycle in which the temperature is gradually reduced until they are unable to improve upon their local solution.

This enables the depths of local minima to be found more quickly, and in tests we have found that this improves the time of convergence to the global solution by orders of magnitude. Additionally, we anneal the ensemble of walkers themselves during the bake-in period, using the temperature schedule proposed in \citet{Hou:2012hd}, and periodically compare the scores of walkers to the best so far, removing those that fall below a pre-determined threshold that depends on the current annealing temperature. After the bake-in period, the algorithm reverts to the vanilla affine-invariant MCMC of \citet{Goodman:2010et} and run for several autocorrelation times, ensuring that detailed balance is maintained.

As inputs to this method, we use the full functional forms of the fallback rate (parameterized as $dM/dt \equiv \md_{\rm sim}$) presented in \citetalias{Guillochon:2013jj} (see their Figure 5) for $\gamma = 4/3$ and $\gamma = 5/3$ polytropes. We assume that as the mass ratio $q \gg 1$, the dependence of $\md$ on $M_{\rm h}$, $M_{\ast}$, and $R_{\ast}$ is self-similar,
\begin{align}
\md &= M_{\rm h,6}^{-1/2} M_{\ast,\odot}^{2} R_{\ast,\odot}^{-3/2} \dot{M}_{\rm int}(\beta),\label{eq:mdotpeak}
\end{align}
where $10^{6} M_{\rm h,6} = M_{\rm h}$, $M_{\odot} M_{\ast,\odot} = M_{\ast}$, $R_{\odot} R_{\ast,\odot} = R_{\ast}$ and $\md_{\rm int}$ is an interpolation of $\md_{\rm sim}$, which was determined in \citetalias{Guillochon:2013jj} directly from $\dmde$ after the debris had relaxed to its final distribution in binding energy $E$ (see Figures 9 and 10 of \citetalias{Guillochon:2013jj}). Note that the \smash{$\md$} functions we use as inputs for our calculation is the rate that the stellar debris returns to pericenter, and not the rate of accretion through the inner edge of the accretion disk. We discuss the validity of this assumption in Section \ref{subsec:dissipative}.

We can eliminate $R_{\ast}$ in Equation \ref{eq:mdotpeak} by using known mass-radius relationships (e.g. \citealt{Tout:1996waa} for MS stars or \citealt{Nauenberg:1972dx} for white dwarfs). For MS stars, we presume that all objects with $M_{\ast} \leq 0.1 M_{\odot}$ have the radius of a $0.1 M_{\ast}$ star. With these relations, $\md$ is solely a function of $M_{\rm h}$, $M_{\ast}$, and $\beta$.

As the simulations of \citetalias{Guillochon:2013jj} are only run for specific values of $\beta$, we determine intermediate $\beta$ solutions by rescaling neighboring simulations in $\beta$-space to the same scaled time variable $x \equiv (t - t_{\min}) / (t_{\max} - t_{\min})$, where $t_{\min}$ and $t_{\max}$ are the minimum and maximum times for each $\md_{\rm sim}$ curve, and then interpolating linearly between the two solutions,
\begin{align}
\lfloor\beta\rfloor &= \min \left\{B\in\beta_{\rm sim} \mid B \geq \beta \right\}\\
\lceil\beta\rceil &= \max \left\{B\in\beta_{\rm sim} \mid B \leq \beta \right\}\\
\dot{M}_{\rm int}(\beta, x) &= \dot{M}_{\rm sim}(\lfloor\beta\rfloor, x)\nonumber\\
&+ \frac{\beta - \lfloor\beta\rfloor}{\lceil\beta\rceil - \lfloor\beta\rfloor} \left[\dot{M}_{\rm sim}(\lceil\beta\rceil, x) - \dot{M}_{\rm sim}(\lfloor\beta\rfloor, x)\right],
\end{align}
where $\beta_{\rm sim}$ is the set of all $\beta$ for which a simulation is available, and where $\lfloor\beta\rfloor$ and $\lceil\beta\rceil$ return the values of $\beta_{\rm sim}$ that bracket $\beta$. We find this preserves the overall shape of the $\md$ curves well for values of $\beta$ for which a simulation is not available, as long as the sampling in $\beta_{\rm sim}$ is sufficiently dense to capture the overall trends.  

The objective function used within \tdefit when comparing our models to the data is the maximum likelihood function,
\begin{equation}
\ln {\cal L}_{\rm LC} = \sum_{i=1}^{j} \left[\frac{\left(V_{{\rm obs},i} - V_{{\rm mod},i}\right)^{2}}{\left(\sigma_{{\rm obs},i}^{2} + \sigma_{\rm v}^{2}\right)}+\ln\left(\sigma_{{\rm obs},i} + \sigma_{\rm v}^{2}\right)\right],
\end{equation}
where $j$ is the number of datapoints, $V_{{\rm obs},i}$ and $V_{{\rm mod},i}$ are respectively the AB magnitude at the $i$th datapoint for the observation and the model, $\sigma_{{\rm obs},i}$ is the measurement error associated with the $i$th datapoint, and $\sigma_{\rm v}$ is the intrinsic variability of the source, assumed to be a constant that scales with black hole mass (see Section \ref{subsec:modeldesc}).

\section{Hydrodynamics of post-disruption debris}\label{sec:debris}
\subsection{Debris stream with self-gravity}\label{subsec:stream}

Determining the fate of the various pieces of the star after a disruptive encounter is critical in determining the appearance of the flare that results from the immense gravitational energy that will be released by the accretion disk that eventually forms. Previously, it has been assumed that self-gravity of the disrupted star is unimportant, and therefore the spread in energy imparted to the debris at pericenter leads to a spread in angle as well as semi-major axis \citep{Strubbe:2009ek,Kasen:2010ci}. Under this assumption, the unbound debris is a homologously expanding structure, which occupies a constant solid angle and whose volume increases proportional to $v_{\rm p}^{3}$.

\begin{figure}
\centering\includegraphics[width=0.85\linewidth,clip=true]{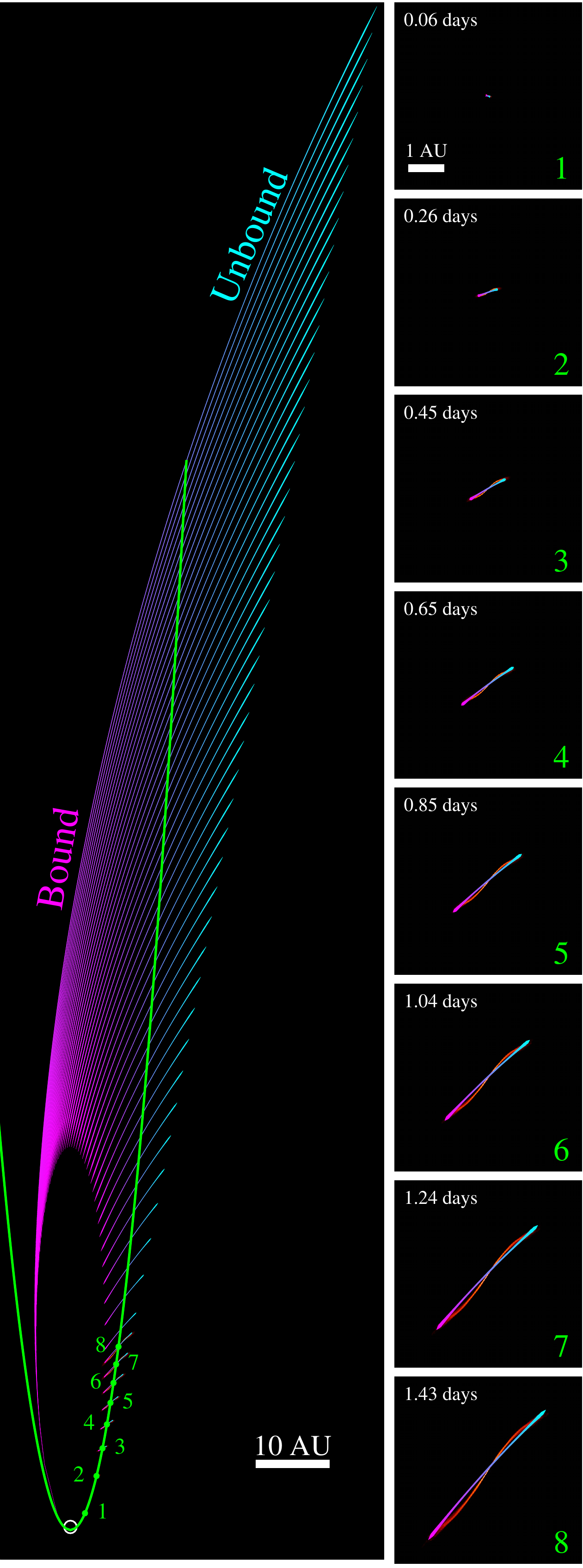}
\caption{Snapshots from a tidal disruption simulation with \smash{$M_{\ast} = M_{\odot}$}, \smash{$M_{\rm h} = 10^{6} M_{\odot}$}, and $\beta = 1.8$, as compared to a simple model of a tidally-confined debris stream with self-gravity, where we assume that the width of the stream scales as $r^{1/4}$ \citep{Kochanek:1994bn}. The left panel shows a superposition of the debris stream at different times, with the longest stream depicting the time when the most bound material returns to pericenter at $t = 42$ days. The color along the stream indicates whether it is bound or unbound from the SMBH, with magenta corresponding to bound and cyan corresponding to unbound. The green line shows the original path of the star, and the green circles show the locations of the surviving core corresponding to the eight snapshots shown on the right-hand side of the figure. In each of the right-hand panels, the simple model of the debris stream is shown atop the results from the simulation.}
\label{fig:thinstreams}
\end{figure}

However, \citet{Kochanek:1994bn} showed that the stream can be in fact gravitationally confined in the transverse direction by self-gravity and forms a very thin structure (Figure \ref{fig:thinstreams}), with a width $\Delta$ and height $H$ that scale as \smash{$\tilde{r}^{1/4}$} for $\gamma = 5/3$, where $\tilde{r} \equiv r/r_{\rm t}$. For general $\gamma$,
\begin{equation}
H^{2} \Lambda^{\frac{2 - \gamma}{\gamma - 1}} \propto {\rm constant}
\end{equation}
where $\Lambda$ is the mass per unit length \citep{Ostriker:1964cg}, which we define to be $\Lambda = M_{\ast}/2R_{\ast}$ at $t = t_{\rm d}$. Assuming that $\Lambda \propto r^{-1}$ \citep{Kochanek:1994bn}, 
\begin{equation}
H = R_{\ast} \tilde{r}^{\frac{2 - \gamma}{2 \gamma - 2}}\label{eq:height},
\end{equation}
where we recover $H \propto \tilde{r}^{1/4}$ for $\gamma = 5/3$.

In the right eight panels of Figure \ref{fig:thinstreams}, we superimpose the results of our hydrodynamical simulations of the disruption of a star by a black hole with mass ratio $q = 10^{6}$ with this simple prescription. We find excellent agreement between the prediction of \citeauthor{Kochanek:1994bn} and our results over the time period in which we ran the simulation. The thinness of the stream is also noted in other hydrodynamical simulations in which the mass ratio is large \citep{Rosswog:2009gg,Hayasaki:2013kd}.

The surface area of this structure for $\gamma = 5/3$ is
\begin{align}
A_{\rm s} &= \frac{2\pi R_{\ast}^{2} q^{1/3}}{\beta} \int_{1}^{r_{\rm u}/r_{\rm p}} \tilde{r}^{1/4} d\tilde{r}\nonumber\\
&\simeq 1.4 \times 10^{-2} M_{6}^{1/3} \beta^{7/8}\left(\frac{t}{t_{\rm ff}}\right)^{5/4} {\rm AU}^{2},
\end{align}
where $r_{\rm u} \simeq r_{\rm t} \beta^{1/2} (t/t_{\rm ff})$ is the distance to which the most unbound material has traveled \citep{Strubbe:2009ek}, and $t_{\rm ff} \equiv \smash{\pi \sqrt{R_{\ast}^{3}/G M_{\ast}}}$ is the star's free-fall time. At the peak time of $\Sim 100$ days, \psone emits $\Sim 10^{45}$ erg s$^{-1}$ of radiation with an effective temperature of a few $10^{4}$ K, implying a photosphere size of $\Sim 10^{15}$ cm with area $\Sim 10^{5}$ AU$^{2}$. By contrast, the area occupied by the stream is only comparable to this value when $t \approx 10^{5} t_{\rm ff} \approx 10$ yr for $q = 10^{6}$ and $\beta = 1$. \citet{Kasen:2010ci} calculated that this component would contribute at most $10^{40}$ ergs s$^{-1}$ of luminosity for the disruption of a solar mass star. However, we believe that this represents an upper limit as the self-gravity of the stream was not included in that work, resulting in an artificially fast rate of recombination.

Because the evolution of the stream is adiabatic, but not incompressible, the stream is resistant to gravitational collapse in both the radial direction perpendicular to the stream, and in the axial direction along the cylinder. Collapse can only occur in the radial direction when $\gamma < 1$, as it becomes energetically favorable to collapse radially \citep{McKee:2007bd}, and can only occur axially for $\gamma > 2$, where the fastest growing mode has a non-zero wavelength and leads to fragmentation \citep[Figure 4 of][]{Ostriker:1964cg,Lee:2007em}.

In a thin stream, the tidal force applied by the black hole results in the density $\rho$ scaling as $r^{-3}$ \citep{Kochanek:1994bn}. As the distance $r \propto t^{2/3}$ when $t \rightarrow \infty$ for parabolic orbits, this implies that $\rho \propto t^{-2}$. For cylinders, the time of free-fall $t_{\rm ff}$ is proportional to $\rho^{-1/2}$, the same as it is for spherical collapse \citep{Chandrasekhar:1961uk}, and thus $t_{\rm ff} \propto t$. Therefore, a segment of the stream within which $t_{\rm ff}$ ever becomes greater than $t$ will not experience recollapse at any future time, as the two timescales differ from one another only by a multiplicative constant. It is also evident that self-gravitating cylinders can be bound whereas a self-bound sphere will not, as the Jeans length progressively decreases in size for structures that are initially confined in fewer directions \citep{Larson:1985to}. This implies that the cylindrical configuration may not remain self-bound if sufficient energy is injected into the star via a particularly deep encounter in which the core itself is violently shocked, which only occurs for $\beta \gtrsim 3$ \citep{Kobayashi:2004kq,Guillochon:2009di,Rosswog:2009gg}, approximately 10\% of disruption events. Additional energy can also be injected by nuclear burning \citep{Carter:1982wf,Rosswog:2008gc}, again only for events in which $\beta$ is significantly larger than 1. For most values of $\beta$, the amount of energy injected into the star at pericenter is insufficient to counteract gravitational confinement in the transverse direction at $t = 0$, and thus from the timescale argument given above the unbound stream would forever be confined.

Given this result, and given the computation burden of resolving a structure with such a large aspect ratio for the number of dynamical timescales necessary for the material to begin accreting onto the black hole, we presume that the gravitational confinement continues to hold at larger distances than we are capable of resolving, and show how the profile of the debris stream would appear if the simulation were followed to the point of the material's return to pericenter in the left panel of Figure \ref{fig:thinstreams}.

By contrast, the bound material travels a much shorter distance from the black hole before turning around. For the material that remains bound to the hole, it has previously been assumed that the material circularizes quickly after returning to pericenter, resulting in an accretion disk with an outer radius equal to $2 r_{\rm p}$, where $r_{\rm p}$ is the pericenter distance \citep{Cannizzo:1990hw,Ulmer:1999jja,Gezari:2009dn,Lodato:2010ic,Strubbe:2011iw}. This is actually a vast underestimate of the distance to which the debris travels, which can be found via Kepler's third law for the orbital period of a body and dividing by two to get the half-period, and then solving for the semi-major axis $a$,
\begin{equation}
r_{\rm o} = 2\left(\frac{G M_{\rm h} t^{2}}{\pi^{2}}\right)^{1/3}\label{eq:ro},
\end{equation}
where we have made the assumption that $r_{\rm o} \simeq 2 a$, appropriate for the highly elliptical orbits of the bound material (The most-bound material has eccentricity $e = 1 - 2q^{-1/3} = 0.98$ for $q = 10^{6}$). As this material is initially confined by its own gravity, the return of the stream to pericenter mimics a huge $\beta$ encounter, which as we explain in the following section can yield a impressive compression ratio.

\begin{figure*}
\centering\includegraphics[width=0.9\linewidth,clip=true]{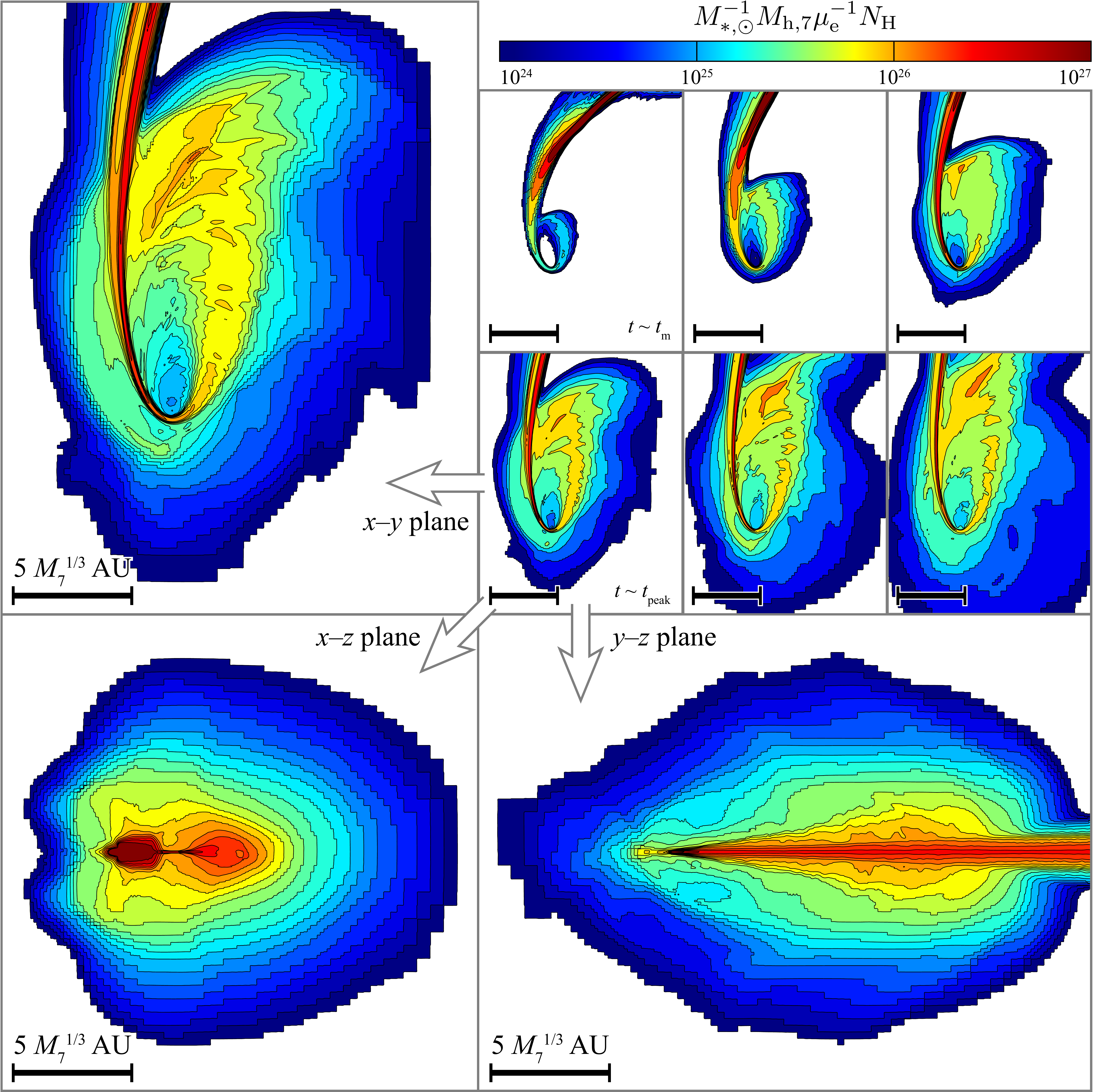}
\caption{Column density contours of the debris resulting from the disruption of a \smash{$M_{\ast} = M_{\odot}$} star by a \smash{$M_{\rm h} = 10^{3} M_{\odot}$} black hole. The column density shown in all panels is scaled to the value that would be expected for a disruption by a \smash{$M_{\rm h} = 10^{7} M_{\odot}$} black hole, with the column density being \smash{$10^{-3.5}$} smaller than what it would be for \smash{$M_{\rm h} = 10^{3} M_{\odot}$}. The six mini-panels in the upper right show the evolution of the column density in the $xy$-plane with time, with the upper left mini-panel showing the column density at the time of return of the most bound material $t_{\rm m}$, and lower left mini-panel showing the column density at the time of peak accretion $t_{\rm peak}$. The three large panels show the column density as viewed from the $x$--$y$, $x$--$z$, and $y$--$z$ planes at $t = t_{\rm peak}$.}
\label{fig:columndensity}
\end{figure*}

\subsection{Dissipative effects within the nozzle}\label{subsec:dissipative}

There are combinations of potentially active mechanisms that can provide the required dissipation for any given event, with each mechanism dominating for particular combinations of $M_{\rm h}$ and $\beta$. To quantify the effect of each of these mechanisms, we define the ratio $\V \equiv (\partial E / \partial t) (T / E)$, where $T$ is the orbital period. $\V$ represents the fraction of gravitational binding energy that is extracted per orbit, with $\V = 1$ indicating a mechanism that fully converts kinetic to internal energy within a single orbit. To have $\dot{M}$ and $L$ trace one another over the duration of a flare, $\V$ must have a value $\gtrsim t_{\rm m} / t_{\rm peak}$, where $t_{\rm peak}$ is the time at which the accretion rate peaks and $t_{\rm m}$ is the time at which the most bound debris returns to pericenter. For partially disruptive encounters, the ratio between these two times is $\sim 3$, but then increases as $\beta^{3}$ for deep encounters in which $t_{\rm m}$ varies more quickly than $t_{\rm peak}$ \citepalias{Guillochon:2013jj}.

In the following sections we provide a brief description of the dissipation mechanisms that are expected to operate in a TDE. Only the first mechanism (hydrodynamical dissipation) is present within our calculations, as we do not include magnetic fields or the effects of a curved space time. Regardless of the origin of the dissipation, we expect that the dissipation observed in our simulations is likely to be quite analogous to the other dissipative process that operate.

\subsubsection{Hydrodynamical dissipation}
As described in Section \ref{subsec:stream}, self-gravity within the debris stream sets its width and height to be equal to $R_{\ast} \tilde{r}^{1/4}$, and thus when the stream crosses the original tidal radius, its height is approximately equal to the size of the original star. If the return of the material to pericenter behaved in the same way as the original encounter, the maximum collapse velocity $v_{\perp}$ would be equal to the sound speed at $r_{\rm t}$ multiplied by $\beta$, yielding a dissipation per orbit ${\V} = q^{-2/3}/\beta$, equal to 2\% for $q = 10^{3}$ and $\beta = 2$ \citep{Carter:1983tz,Stone:2013gk}. In our hydrodynamical simulations for $q = 10^{3}$, we find that $\approx 10\%$ of the total kinetic energy of the debris is dissipated upon its return to pericenter through strong compression at the nozzle (Figure \ref{fig:columndensity}). This is a factor of a few larger than the expected dissipation.

However, as the star has been stretched tremendously, the sound speed within the stream has dropped by a significant factor, meaning that the distance from the black hole at which the stream's sound-crossing time is comparable to the orbital time (i.e., where the tidal and pressure forces are approximately in balance) is not the star's original tidal radius, but is instead somewhat further away.

For two points that are separated by a distance $dr$ within the original star, their new distance $dr^{\prime}$ upon returning to pericenter is related to the difference in binding energy between them, which remains constant after the encounter and is equal to
\begin{equation}
\frac{dE}{dr} \simeq \frac{E(r_{\rm p})}{r_{\rm p}} = \frac{G M_\ast}{R_{\ast}^{2}} q^{1/3}\label{eq:deg}
\end{equation}
As angular momentum is approximately conserved, the two points will cross pericenter at the same location they originally crossed pericenter, but at two different times $t$ separated by a time $dt$ owing to their different orbital energies. Assuming that the star originally had approached on a parabolic orbit, and that all the bound debris are on highly elliptical orbits, the distance from the black hole is
\begin{equation}
r' = \left(\frac{9}{2} G M_{\rm h} t^{2}\right)^{1/3},\label{eq:rprime}
\end{equation}
and thus
\begin{equation}
\frac{dr^{\prime}}{dt} = \left(\frac{4 G M_{\rm h}}{3 t}\right)^{1/3}.\label{eq:drprimedt}
\end{equation}
If we set $t = 0$ to be the time when the first point re-crosses pericenter, the difference in time is simply the difference in orbital period, and thus we can use Kepler's third law to derive $dE/dt$. Using Equations \ref{eq:deg} and \ref{eq:drprimedt}, we can use the chain rule to determine $dr^{\prime}/dr$,
\begin{equation}
\frac{dr^{\prime}}{dr} = \frac{dE}{dr}\frac{dt}{dE}\frac{dr^{\prime}}{dt} = \left(\frac{3\pi}{2}\right)^{2/3} \left(\frac{t}{t_{\rm m}}\right)^{4/3} q^{2/3}.\label{eq:drdrprime}
\end{equation}
For $q = 10^{3}$, this implies that the material has been stretched by a factor $\gtrsim 10^{2}$ after the most-bound material begins accreting, and for $q = 10^{6}$ this factor is $\gtrsim 10^{4}$. The change in volume is then given by the change in cross-section of the stream multiplied by the change in length given by Equation \ref{eq:drdrprime},
\begin{equation}
\frac{dV^{\prime}}{dV} = \frac{dr^{\prime}}{dr} \tilde{r}^{\frac{2-\gamma}{\gamma-1}}
\end{equation}
where we have used Equation \ref{eq:height} to estimate the width and height of the stream.

The density of the stream $\rho$ as it returns to pericenter can be approximated by assuming that $dM/dr = \Lambda$, although in reality this distribution can be determined more exactly from the numerical determination of $\md$ by a change of variables from $E$ to $r$. Under this assumption, the change in density is simply related to the change in volume alone. As the tidal radius is proportional to $\rho^{1/3}$, the ratio of the effective tidal radius of the stream $r_{\rm t,s}$ to $r_{\rm t}$ is then
\begin{align}
\frac{r_{\rm t,s}}{r_{\rm t}} &= \left(\frac{dV^{\prime}}{dV}\right)^{1/3}\nonumber\\
&= \left(\frac{3\pi}{2} q \right)^{\frac{2}{3}\frac{\gamma-1}{4\gamma-5}} \left(\frac{t}{t_{\rm m}}\right)^{\frac{4}{3}\frac{\gamma-1}{4\gamma-5}} \equiv \frac{\beta_{\rm s}}{\beta}\label{eq:rtstream}
\end{align}
where we have substituted $r_{\rm t,s}/r_{\rm t}$ for $\tilde{r}$, and where we have presumed that the time until the stream reaches pericenter from $r_{\rm t,s}$ is small compared to the time since disruption.

Under the assumption that the stream expands adiabatically and its pressure is governed by ideal gas pressure, this results in a reduction in the sound speed $c_{\rm s} = \sqrt{dP/d\rho} \propto V^{(1 - \gamma)/2}$, where $\gamma$ is the adiabatic index of the fluid. At $r = r_{\rm t,s}$, the ratio of the sound speed within the stream $c_{\rm s,s}$ to the star's original sound speed $c_{\rm s,\ast}$ is
\begin{align}
\frac{c_{\rm s,s}}{c_{\rm s,\ast}} &= \left(\frac{dV^{\prime}}{dV}\right)^{\frac{1 - \gamma}{2}}\nonumber\\
&= \left(\frac{3 \pi}{2} q\right)^{\frac{(\gamma-1)^{2}}{5-4\gamma}} \left(\frac{t}{t_{\rm m}}\right)^{\frac{2(\gamma-1)^{2}}{5-4\gamma}}\label{eq:csstream}.
\end{align}

Analogous to the original star, the maximum collapse velocity of the stream $v_{\perp, {\rm s}}$ is equal to the sound speed at $r_{\rm t,s}$ multiplied by the stream's impact parameter $\beta_{\rm s} \equiv \beta r_{\rm t,s} / r_{\rm t}$, $v_{\perp} = \beta_{\rm s} c_{\rm s,s}$. As the majority of the dissipation comes through the conversion of the kinetic energy of the vertical collapse via shocks, the fractional change in the specific internal energy is
\begin{align}
\V_{\rm H} &= \frac{\beta_{\rm s}^{2} c_{\rm s,s}^{2}}{v_{\rm p}^{2} }\nonumber\\
&= \left(\frac{3\pi}{2} q\right)^{-\frac{2}{3}\frac{(\gamma - 1)(3\gamma-5)}{4\gamma-5}} \left(\frac{t}{t_{\rm m}}\right)^{-\frac{4}{3}\frac{(\gamma - 1)(3\gamma-5)}{4\gamma-5}} \beta q^{-2/3}.\label{eq:vhydro}
\end{align}

As $r_{\rm t} \propto \rho^{1/3}$, and $c_{\rm s} \propto \rho^{-1/3}$ for $\gamma = 5/3$, Equations \ref{eq:rtstream} and \ref{eq:csstream} are inverses of one another in the adiabatic case,
\begin{equation}
\frac{r_{\rm t,s}}{r_{\rm t}} = \frac{c_{\rm s,\ast}}{c_{\rm s,s}} = 60 M_{6}^{4/15} \left(\frac{t}{t_{\rm m}}\right)^{8/15}
\end{equation}
and thus \ref{eq:vhydro} simplifies to $\V_{\rm H} = \beta q^{-2/3}$, identical to the amount of dissipation experienced by the original star. One key difference exists between the original encounter and the stream's return to pericenter: While the original encounter may only result in the partial shock-heating of the star, even for relatively deep $\beta$ \citep{Guillochon:2009di}, the fact that the collapse of the stream is highly supersonic ($\beta_{\rm s} \sim 60 \beta$) guarantees that shock-heating will occur upon the material's return to pericenter.

\begin{figure}
\centering\includegraphics[width=\linewidth,clip=true]{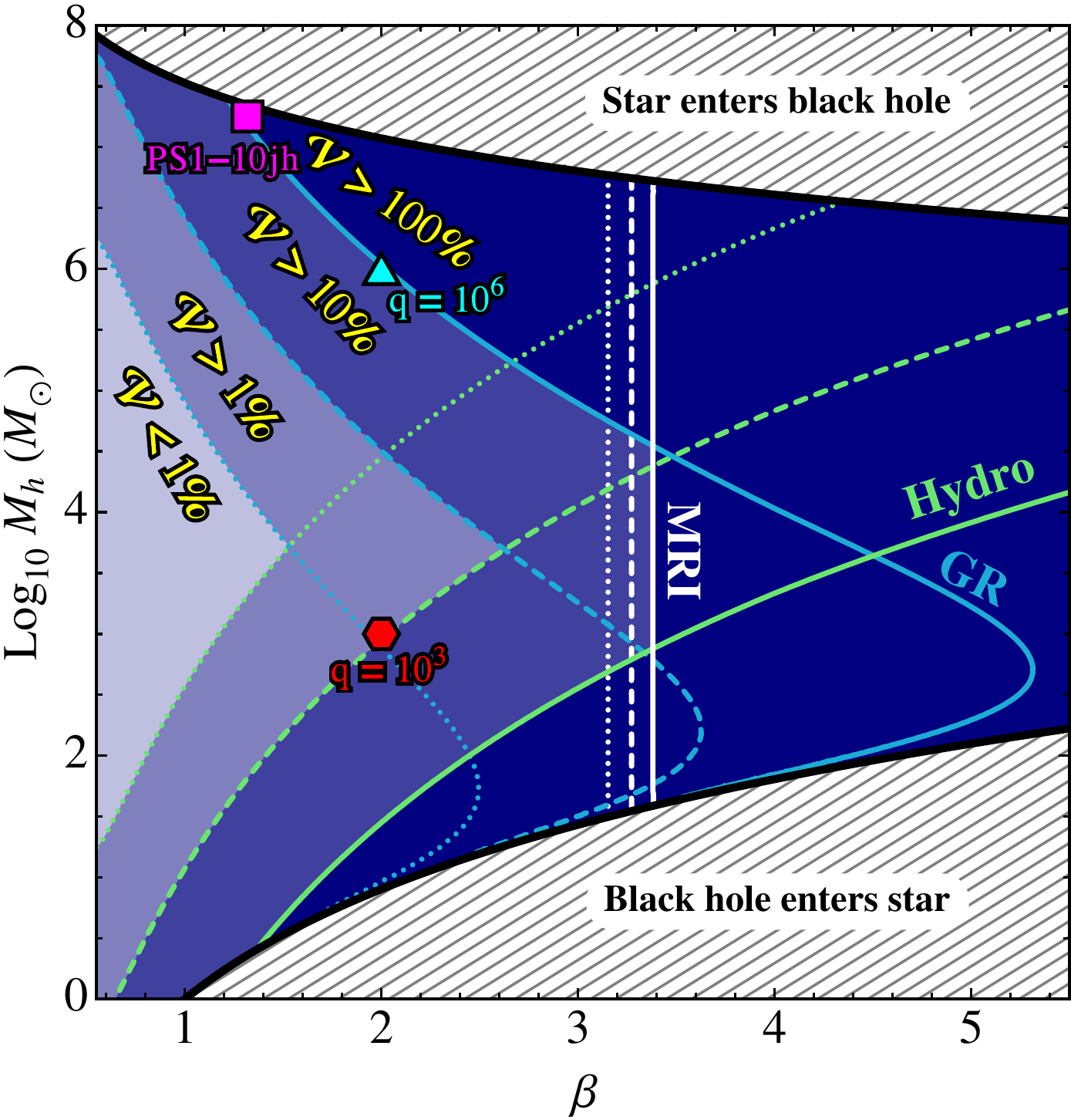}
\caption{Fraction of binding energy dissipated at $t = t_{\rm peak}$ for three mechanisms that may contribute to the circularization of material after a tidal disruption, with mint corresponding to hydrodynamical shocks at the nozzle point, light blue corresponding to dissipation through GR precession (presuming $\gamma = 5/3$), and white corresponding to the MRI mechanism. For each mechanism, three contours of $\V$ are shown, with solid corresponding to 100\%, dashed corresponding to 10\%, and dotted corresponding to 1\%. If all three mechanisms operate, the shaded blue regions represent zones in which $\V$ adopts the values specified by the unions of the regions enclosed by the three sets of contours, with the lightest/darkest corresponding to the least/most dissipation. For reference, our two hydrodynamical simulations are shown by the cyan triangle and the red hexagon, and the highest-likelihood fit returned by our MLA (Section \ref{subsec:generalmodel}) is shown by the magenta square.}
\label{fig:dissipation}
\end{figure}

For our $q = 10^{3}$ simulation, the amount of dissipation expected per orbit predicted by Equation \ref{eq:vhydro} is $4 \times 10^{-2}$, and for our $q = 10^{6}$ simulation the expected dissipation would only be $4 \times 10^{-4}$. Thus, the conversion of kinetic energy to internal energy via shocks at the nozzle point is inefficient for all but the lowest mass ratios and/or the largest impact parameters, and would be incapable of circularizing material on a timescale that is shorter than the peak timescale of \psone (Figure \ref{fig:dissipation}). This suggests that a viscous mechanism that involves an unresolved hydrodynamical instability, or a mechanism that is beyond pure hydrodynamics, is responsible for the circularization of the material for this event.

An additional complication that is not addressed here is recombination. As the stream expands, its internal temperature drops below the point at which hydrogen begins to recombine, flooring its temperature to $\Sim 10^{4}$ K until most of the hydrogen is neutral \citep{Roos:1992cj,Kochanek:1994bn}. This implies that the ratio between the initial and final sound speeds is somewhat smaller than when assuming adiabaticity holds to arbitrarily-low stream densities, and depends on the initial temperature of the fluid, which is $\sim 10^{4}$ K in the outer layers of the Sun, but $\Sim 10^{7}$ K in its core. This also causes the stream to expand somewhat due to the release of latent heat. However, as the material returns to pericenter, the compression of the material will reionize it. Given these complications, it is unclear if this process would lead to more or less dissipation at the nozzle.

\subsubsection{Dissipation through General Relativistic Precession}

For orbits in which the pericenter is comparable to the Schwarzschild radius $r_{\rm g}$, the orbital trajectory begins to deviate from elliptical due to precession induced by the curved space-time. The precession time in the inner part of the disk is \citep{Valsecchi:2012gw}:
\begin{equation}
\dot{\gamma}_{\rm GR} = \left(\frac{2\pi}{T}\right)^{5/3} \frac{3 G^{2/3}}{c^{2}} \frac{M_{\rm h}^{2/3}}{1-e^{2}},\label{eq:prec}
\end{equation}
where $T$ and $e$ are respectively the period and eccentricity of the stream. As the debris resulting from a tidal disruption has a range of pericenter distances ($r_{\rm p} \pm R_{\ast}$), there is a gradient in precession times of the returning debris. This precession causes the orbits to cross one another, dissipating energy \citep{Eracleous:1995fv}. When compared to the standard $\alpha$ viscosity prescription, the timescale of this precession is comparable to the viscous time,
\begin{equation}
t_{\rm prec} = 10^{-1.7} M_{6} T_{5}^{-1/2}\left(1+e\right)\left(\frac{r_{\rm p}}{r_{\rm g}}\right)^{2}\;{\rm yr},
\end{equation}
where $T_{5} \equiv 10^{5} T$ is the local disk temperature. In a tidal disruption, the most-bound material is also the material with the shortest precession time, and it is this timescale that sets the overall rate of dissipation. By setting $T$ and $e$ in Equation \ref{eq:prec} to $t_{\rm m} \equiv \sqrt{q/2} \beta^{-3} t_{\rm ff}$ and $e_{\rm m} \equiv 1 - 2 \beta q^{-1/3}$, the period and eccentricity of the most-bound material, and by assuming that precession through an angle $2 \pi$ would lead to complete dissipation, the dissipation due to relativistic precession for the material that corresponds to the peak in the accretion rate is
\begin{equation}
\V_{\rm GR, peak} = \frac{t_{\rm peak}}{t_{\rm m}} \left(\frac{2\pi}{t_{\rm m}}\right)^{2/3} \frac{3 G^{2/3}}{c^{2}} \frac{M_{\rm h}^{2/3}}{1-e_{\rm m}^{2}}.\label{eq:vgr}
\end{equation}

In general, the period of the most-bound material tends to smaller values for larger $q$ and $\beta$, resulting in more dissipation, except in the case that $r_{\rm p}$ and $R_{\ast}$ are comparable (Figure \ref{fig:dissipation}, cyan curves). If none of the other dissipation mechanisms are effective, this means that disruptions by massive black holes, in which $r_{\rm p}$ and $r_{\rm g}$ are closer to one another in value, would be the only cases in which $\md$ and $L$ follow one another closely. As we will show in Section \ref{sec:ps1-10jh}, the highest-likelihood models of \psone seem to be consistent with a relativistic encounter with the central black hole, so it is possible that \psone's identification as a TDE was contingent upon the condition that $r_{\rm p} \sim r_{\rm g}$.

\subsubsection{Super-Keplerian, Compressive MRI}

The magnetic field in the bound stellar debris is likely to be amplified by compression and by strong shearing within the nozzle region. The effects of magnetic shearing, the magneto-rotational instability \citep[MRI;][]{Balbus:1998tw}, is expected to lead to the rapid exponential growth of the magnetic field with a characteristic timescale of order the rotational period. This instability has been routinely studied in the context of accretion disks and we argue here that it is likely to operate within the nozzle. However, given that the fluid's motion is super-Keplerian at pericenter (being on near-parabolic orbits), its exact character is difficult to compare directly to the classical MRI, as the boundary conditions are constantly changing and the material is never in steady-state.

The MRI is present in a weakly magnetized, rotating fluid wherever
\begin{equation}
{d\Omega^2 \over d \ln r}<0.
\end{equation}
The ensuing growth of the field is exponential with a characteristic  time scale given by  $t_{\rm MRI}=  4\pi |d\Omega/d\ln r|^{-1} $ \citep{Balbus:1998tw}. For a (super-)Keplerian angular velocity distribution $\Omega \propto r^{-3/2}$, this gives $t_{\rm MRI}=(4/3)\Omega^{-1}$.  Exponential growth of the field on the timescale $\Omega^{-1}$ by the MRI is likely to dominate over other amplification process such as field compression within the same characteristic time. While a variety of accretion efficiencies are reported in numerical realizations of magnetically-driven accretion disks, which depend on the geometry, dimensionality of the simulation, and included physics, essentially all models find that the strength of the magnetic field amplifies to the point that it is capable of converting fluid motion into internal energy. From global simulations of the MRI, the build-up of the magnetic field strength is confirmed to be exponential, resulting in a time to complete saturation being a constant multiple of the orbital period. In \citet{Stone:1996hk}, this constant is found to be 3.  Once the magnetic field strength is saturated, the resulting angular momentum transport  will be governed by the turbulence and is therefore expected to take place over longer timescales. A simple estimate of the saturation field can be obtained by equating the characteristic mode scale, $\sim v_{\rm A} (d\Omega/d\ln r)^{-1}$, where $v_{\rm A}$ is the Alfv\'{e}n velocity, to the shearing length scale, $\sim dr/d\ln \Omega$, such that \smash{$B_{\rm sat} \sim (4\pi \rho)^{1/2} \Omega r$}. This saturation field is achieved after turbulence is fully developed, which in numerical simulations takes about a few tens of  rotations following the initial exponential growth \citep{Hawley:1996gh,Stone:1996hk}.
 
For the Sun, the initial interior magnetic field energy at the base of the convective zone $E_{B,0} \sim 10^{-10} E_{\rm g}$ \citep{Miesch:2009kb}, although larger initial fields are possible in general \citep{Durney:1993hq}. As the tidal forces stretch the star into a long stream, the volume of the fluid increases by a factor $\beta_{\rm s}^{3}$ (Equation \ref{eq:rtstream}) prior to returning to pericenter, reducing the magnetic field strength further. However, when the stream returns to pericenter, it experiences a dramatic decrease in volume by a factor $\smash{\beta_{\rm s}^{2/(\gamma-1)}}$ \citep{Luminet:1986ch}. Assuming the frozen flux approximation, the new magnetic energy density is
\begin{align}
E_{B} &= E_{B,0} \beta_{\rm s}^{\frac{3\gamma - 5}{\gamma - 1}}\\
&= \left(\frac{3\pi}{2} q \right)^{\frac{2}{3}\frac{3\gamma - 5}{4\gamma - 5}} \left(\frac{t}{t_{\rm m}}\right)^{\frac{4}{3}\frac{3\gamma - 5}{4\gamma - 5}}
\end{align}
For $\gamma = 5/3$, the dependence on $\beta_{\rm s}$ disappears, i.e. the magnetic field strength upon return to pericenter is identical to the star's initial interior field. Assuming that the magnetic field within the debris has strength $E_{B}$ relative to the local gravitational binding energy $E_{\rm g}$ upon returning to the nozzle, $\V_{\rm MRI}$ adopts a simple form (Figure \ref{fig:dissipation}, white curves),
\begin{equation}
\V_{\rm MRI, peak} = \frac{E_{B}}{E_{\rm g}} \exp \left[\frac{t_{\rm peak}}{3 t_{\rm m}}\right]\label{eq:vmri}.
\end{equation}

\subsection{Is the Debris Disk Dissipative Enough?}\label{subsec:disenough}
In order for the emergent luminosity $L$ to follow the feeding rate $\md$ closely, the dissipation must be effective enough such that material returning to pericenter can circularize on a timescale $t_{\rm c}$ that is at most the time since disruption $t_{\rm d}$.

In the original calculation of \citet{Cannizzo:1990hw}, the initial conditions place a fixed amount of a mass a fixed distance from the black hole at $t = 0$. The matter is then allowed to evolve viscously, resulting in the transport of mass inwards, and the transfer of angular momentum outwards. While this initial condition is acceptable for fallback calculations onto a newly-formed neutron star in a long GRB \citep{Lee:2006kp,Kumar:2008fa,Cannizzo:2009gm,LopezCamara:2009iq,Milosavljevic:2012ci} and for a compact binary merger \citep{Lee:2004fn,Metzger:2008hp,Lee:2009iy,Metzger:2009fu}, it is likely not a perfect analogue for a tidal disruption of a star originally on a parabolic orbit, as it neglects the continuous injection of energy from the returning stream.

As material circularizes, it must deposit $(\sqrt{2} - 1)^2 v_{\rm K}^2$ of kinetic energy within a few orbital periods, as it has to slow down from a near-escape velocity to the local Keplerian velocity $v_{\rm K}$. If the circularization is rapid, this additional source of heating leads to an $H/R \sim 1$ and consequently rapid accretion. For rapid accretion, the accretion disk mass remains small, on order $\dot{M}(t) t_{\rm c}$. The total mass accreting onto the black hole via the stream at any one time samples a later segment of the fallback curve, offset by a time $t_{\rm c}$,  $\dot{M}(t + t_{\rm c}) t_{\rm c}$. At early times, this mass is always larger than the amount of mass in the disk, as $\dot{M}$ is increasing rapidly with time. This is very similar to the argument made below equation 34 in \citet{Kumar:2008fa} for the direct fallback phase of a GRB. At late times, this mass a bit less than the mass in the disk, but not considerably so as $t_{\rm c} \ll t_{\rm fb}$.

So long as the returning material in the debris stream has a comparable mass to the mass present in the accretion disk and circularization is rapid, the timescale for accretion can remain short over the full duration of the flare. If however the returning debris is unable to circularize quickly at some point in the flare's evolution, matter will build up in a disk with $H/R \ll 1$, with the resulting disk mass being somewhat larger than the incoming mass. This would effectively ``erase'' the incoming $\dot{M}$'s functional form, and instead result in a fallback rate with a much shallower slope, between $-1$ and $-4/3$ \citep{Cannizzo:1990hw}. Evidence for such a transition may have been seen in Swift-J1644 \citep{Cannizzo:2011df}. However, the time of the transition is likely to occur at very late times for an encounter with $\beta \sim 1$, as suggested by our most-likely solutions (see Section \ref{subsec:generalmodel}). Using Equation 21 from \citet{Cannizzo:2011df} we find that this transition would occur at $\sim 10^{3}$ yr for these parameters, well beyond the time at which \psone's flux dropped below that of its host galaxy.

In Figure \ref{fig:dissipation} we show the three sources of dissipation that we estimated above. We find that while significant dissipation is expected for large $\beta$ encounters, or for encounters in which $r_{\rm p} \sim r_{\rm g}$ (as is the case for our best-fitting model for \psone), that there are many combinations of $\beta$ and $q$ that may not have the required dissipation necessary to ensure the direct mapping in time of $\md$ to $L$. In our $q = 10^{3}$ hydrodynamical simulation, we found somewhat more dissipation than what is expected from a simple analytical calculation, but the resolution at which we resolved the compression at pericenter was only marginally sufficient to resolve the strong shocks that form there.

One potential resolution to this issue is the adiabatic index of the fluid $\gamma$, which in the above calculations we have assumed $ = 5/3$, although the real equation of state within the stream is likely softer due to the influence of recombination. With a softer equation of state, the cancelations that occur for $\gamma = 5/3$ and eliminate the dependence on $\beta_{\rm s}$ for the hydrodynamical (Equation \ref{eq:vhydro}) and MRI (Equation \ref{eq:vmri}) dissipation mechanisms would no longer apply, yielding both increased compression and magnetic field strengths, and thus additional dissipation.

While the initial dissipation of the stream may indeed come as the combination of the three previously described mechanisms, it is likely that the mechanism responsible for the accretion onto the black hole once the material has been assembled into a disk is the MRI mechanism, as is suspected for steadily-accreting AGN. Given the computational challenge of simultaneously resolving the nozzle region and the full debris stream, it is clear that local high-resolution magnetohydrodynamic simulations are required to determine the true dissipation rate $\V$ at the nozzle.

\section{The Relationship Between Steadily-Accreting AGN and TDE Debris Disks}\label{sec:steady}
Within the debris structure formed from a tidal disruption, the same mechanisms that operate in steady-state AGN may continue to operate. There are a number of differences between the structure of a debris disk resulting from tidal disruption and the structure of steadily-accreting AGN, but we will argue that similar processes are responsible for the appearance of both structures. In this section, we will make continued reference to the highest-likelihood model of \psone, which is determined in Section \ref{sec:model}.

\subsection{The Conversion of Mass to Light}
For steadily-accreting AGN, energy is thought to be released by the viscous MRI process at all radii. The amount of energy available at a particular radius depends on the local gravitational potential, and thus the vast majority of the energy emitted by accreting black holes is produced within a few times the Schwarzschild radius $r_{\rm g}$. The temperature profile that results from this release in energy within the accretion disk is given by the well-known expression first presented in \citet{LyndenBell:1969dq}, and scales as $r^{-3/4}$, resulting in a sum of blackbodies with a continuum slope $F_{\nu} \propto \nu^{1/3}$ \citep{Pringle:1972vb,Gaskell:2008wx}. AGN are divided into two fundamental categories \citep{Antonucci:2012vl}: Compton-thick (i.e. non-thermal) AGN, which are obscured by $\Sim 10^{24}$ cm$^{2}$ of column and thus making them optically thick to Compton scattering \citep{Treister:2009bk}, and thermal AGN, which have column densities significantly less than this value, enabling the black hole's emission to be directly observed. However, all thermal AGN show an excess in the blue known as the ``big blue bump'' \citep{Shields:1978ce,Lawrence:2012hn}, and the slopes of their continuum $F_{\nu} \propto \nu^{-1}$ \citep{Gaskell:2009dc}. This is more consistent with the notion that the light emitted from the central parts of the disk is intercepted by intervening gas before it is observed, with nearly one hundred percent of the light emitted by the disk being reprocessed in this way. This implies that a significant fraction of the mass that may eventually be accreted by the black hole is suspended some distance above the disk plane, where it can intercept a large fraction of the outgoing light.

\begin{figure*}
\centering\includegraphics[width=0.9\linewidth,clip=true]{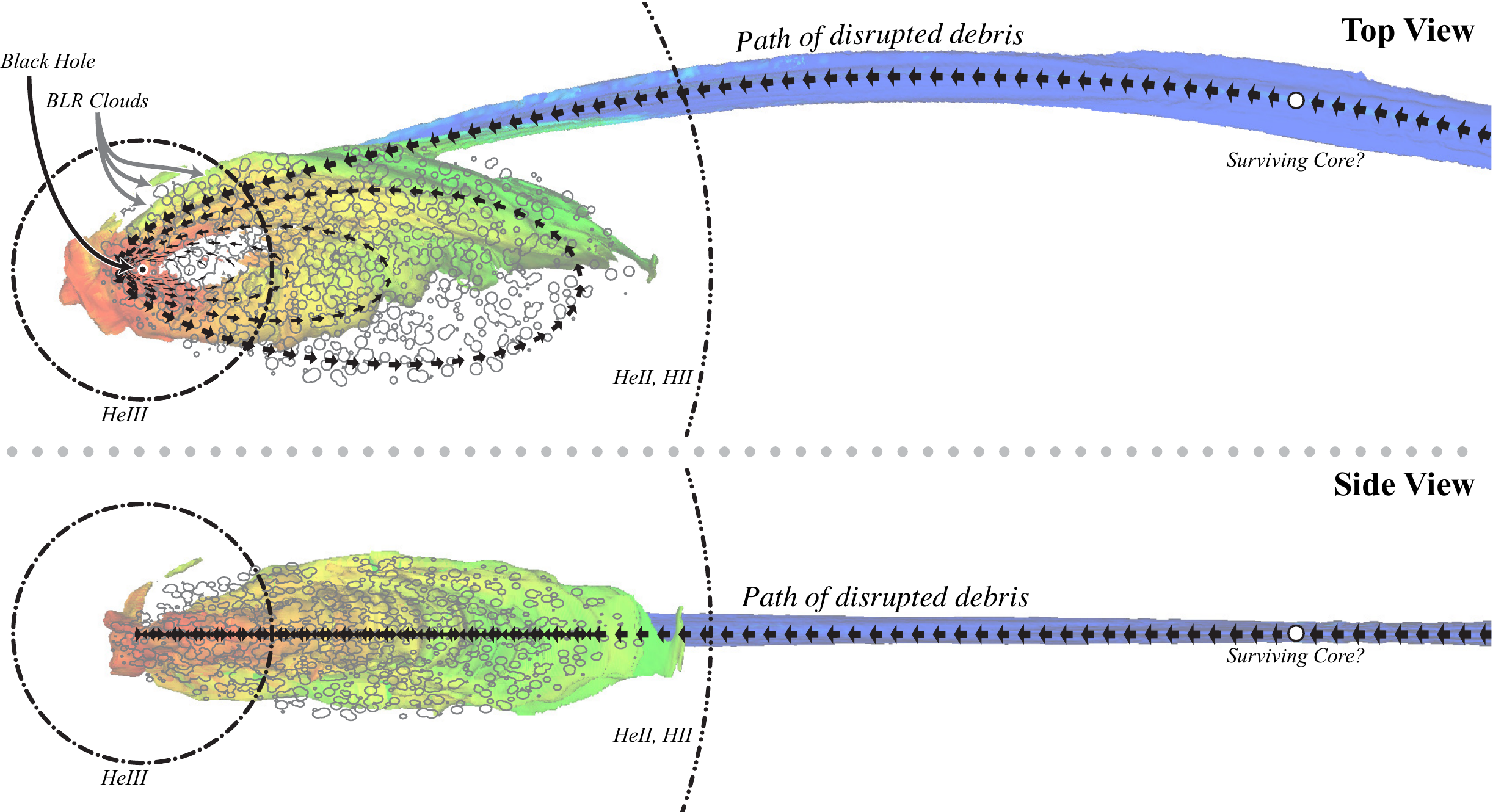}
\caption{Schematic figure from our $q = 10^{3}$ simulation demonstrating the geometry of the debris resulting from a tidal disruption at $t - t_{\rm d} = 4.3 \times 10^{5}$ s, shown from the top (top image) and the side (bottom image). The three-dimensional isodensity contours are colored according their temperature, with red being hot and blue being cold. Super-imposed on these contours are a line of arrows showing the path of the circularizing debris as return to the black hole (black disk), inside which a surviving stellar core may reside (white disk). The dot-dashed and double-dot-dashed lines respectively show the regions interior to which helium is doubly-ionized and hydrogen/helium are singly ionized within the BLR. The region in which BLR clouds may form is super-imposed using the gray contours, although we note that the BLR may instead be in the form of a diffuse wind.}
\label{fig:schematic}
\end{figure*}

For the accretion structure that forms from the debris of a tidal disruption, the dissipation at the nozzle point provides a means for lifting material above and below the orbital plane, resulting in a sheath of material that surrounds the debris and is very optically thick for certain lines of sight (Figure \ref{fig:columndensity}). However, as the spread in energy at the nozzle point does not completely virialize the flow, the resulting distribution of matter is flattened, allowing the central regions of the accretion disk to be visible through material that is close to the Compton-thick limit. The time-series presented in the upper six panels of Figure \ref{fig:columndensity} show that despite the continually active dissipation process at the nozzle-point, the region directly above the central parts of the accretion disk remain relatively evacuated of gas before, during, and after the time of peak accretion. For the toy simulation presented here, the optical depth to Thomson scattering directly above the black hole and perpendicular to the orbital plane of the debris is $\sim 1$, depending on the electron fraction of debris (assumed to be pure hydrogen in Figure \ref{fig:columndensity}). For more massive black holes, the debris is spread over a larger volume, as the tidal radius grows as \smash{$M_{\rm h}^{1/3}$}, and thus \smash{$N \propto M_{\rm h}^{-1}$}. If the dissipation rate were the same independent of black hole mass, it would be expected that the disruption of stars by more massive black holes would yield more lines of sight for which $\tau \sim 1$.

\subsection{Source of Broad Emission Lines}\label{subsec:blrsource}

Broad line emission is visible in many AGN, being thought to be produced by gas above and below the disk plane at distances from light hours to light years away from the black hole. For other AGN, this region is not directly observable, which has been attributed to a torus at large radii that can obscure the broad line region for lines of sight that run within a few tens of degrees of the disk plane \citep[i.e. the AGN unification model,][]{Antonucci:1993fe}. The emission lines produced within this region have been successfully used to measure black hole masses \citep{Dibai:1977tr,Peterson:2004ig,Marziani:2012gj} based on measurements of the time lag in the response of line luminosity to variations in the output of the central engine \citep[see e.g.][]{Denney:2009gm}. It is still debated whether this material is in the form of an optically-thick disk wind \citep{Trump:2011bp} or optically-thick clouds \citep{Celotti:1999tm}, but in either case the material that constitutes the BLR is mostly bound to the black hole \citep{Proga:2008bz,Pancoast:2012dd}.

In a steady-accreting AGN, material accretes from very large distances (\smash{$\gtrsim 10^{3} r_{\rm g}$}), and the emission from this region is often manifest as an IR bump in Type II AGN \citep{Koratkar:1999di}. At such distances, the ionizing flux originating from the black hole is not sufficient to maintain a large ion fraction within the disk's emitting layer. The closer one gets to the central black hole, the greater the incident flux of ionizing radiation on the BLR wind/clouds that generate the observed emission lines. The fraction of atoms in an excited state X$_{+}$ relative to the state directly below it X$_{0}$ is approximately \citep{Osterbrock:2006ul}
\begin{equation}
\frac{n_{{\rm X}_{+}}}{n_{{\rm X}_{0}}} \sim \frac{a(\nu_{\rm ion})}{\alpha_{\rm B}({\rm X}_{0}) h \nu_{\rm ion}}\frac{Q({\rm X}_{0})}{4\pi r^{2} n_{\rm e}},
\end{equation}
where $Q({\rm X}_{0})$ is the flux in photons capable of ionizing the lower state. This expression shows that that as the distance from the central engine increases, the number of atoms in the high state decreases, assuming that the electron density $n_{\rm e}$ decreases with radius more slowly than $r^{-4}$ (as $Q \propto r^{-2})$, and also shows that species with larger ionization potentials will have less atoms in the high state than species with smaller ionization potentials. This leads to a hierarchy of ions in the disk, with those with the highest ionization potential being predominant in the inner regions of the disk. In a steadily-accreting AGN, the flux in ionizing photons is large enough to fully ionize iron \citep[as evidenced by the existence of Fe K lines,][]{Fabian:2000hr}, and given that atoms at large radii are mostly neutral, all ionic species of all elements exist at some distance from the central engine. Reverberation mapping supports this basic photoionization picture, as $R_{\rm BLR} \sim L^{1/2}$ \citep{Bentz:2010hk,Bentz:2013hm}. In particular, the optical wave band hosts several lines from the Balmer series of hydrogen and lines from both singly-ionized and neutral helium \citep{Bentz:2010hk}.

This wide range of scales is in stark contrast to the debris disk formed as the result of a tidal disruption, which we schematically illustrate in Figure \ref{fig:schematic}. Rather than material spiraling in from parsec scales, material is instead ejected from the nozzle point, which lies at the star's original point of closest approach, and typically has scales on the order of a few AU. As a result, the debris disk forms from the inside out. The ratio of line strengths in a TDE is dependent upon the number of atoms in the photosphere that are in the particular ionization state associated with each line. For \psone, the lack of an \ha emission line was interpreted by \citetalias{Gezari:2012fk} as being attributed to a lack of hydrogen atoms. However, the Balmer series requires neutral hydrogen to be present in sufficient quantities to produce a line in excess of the continuum emission. As shown in our $q = 10^{3}$ simulation, material is ejected from the nozzle point at approximately the escape velocity, with the fastest moving material traversing a distance $r_{\rm t} [(t-t_{\rm d})/t_{\rm p}]^{2/3}$. {\it This sets an upper limit on the radial extent of the disk}. Therefore, the lack of an observed emission feature may simply be the result of the disk not being large enough to host the region required for that particular feature's production.

\begin{figure}
\centering\includegraphics[width=\linewidth,clip=true]{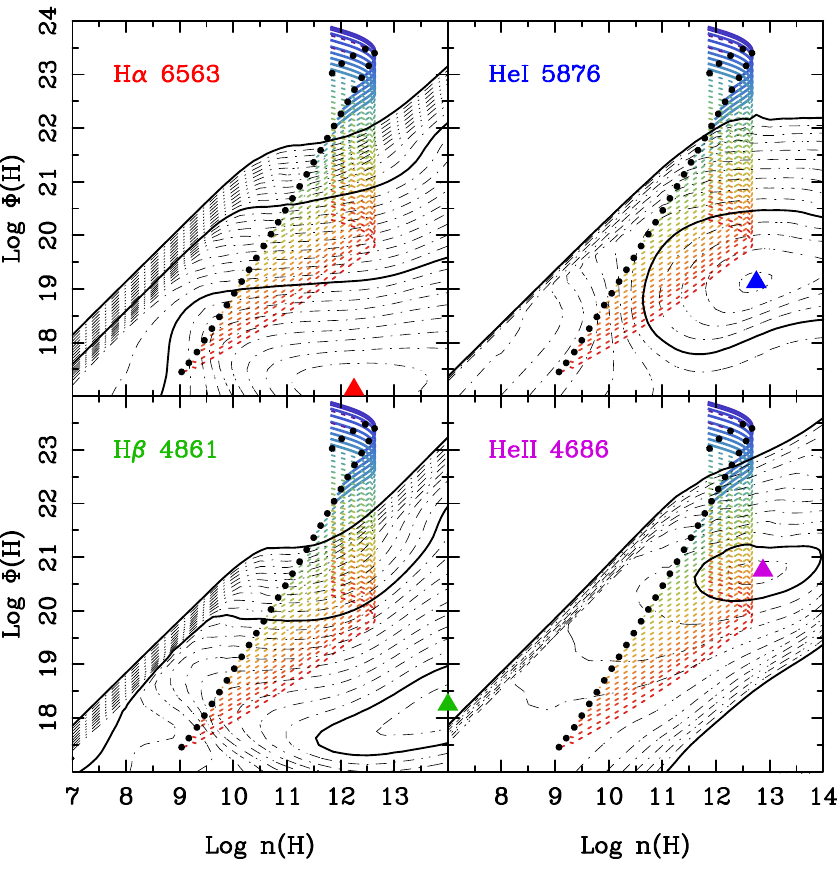}
\caption{Contours of the log of the equivalent width of four emission lines as a function of hydrogen--ionizing flux $\Phi({\rm H})$ and hydrogen number density $n({\rm H})$, where the black dashed and solid contours correspond to 0.1 and 1 decade, respectively \citep[Adapted from Figure 1 of][]{Korista:2004cp}, with the smallest contour corresponding to \smash{1 \AA} of equivalent width. The colored triangle within each panel indicates the peak equivalent width for each line. The rainbow-colored curves show the profiles of the debris in the $\Phi({\rm H})-n({\rm H})$ plane resulting from the tidal disruption that corresponds to the highest-likelihood fit of \psone (Section \ref{subsec:generalmodel}). The curves span the full duration of the event, with the solid curves corresponding to the range between the first and last spectrum taken for the event (with purple being -22 and red being +358 days from peak), and the dashed curves corresponding to unobserved epochs before/after the spectral coverage. The black dotted curve shows the conditions at $r_{\rm o}$ as a function of time over the full event duration. Note that \ha, \hb, and \hei would potentially be observable if additional spectra were collected at later times.}
\label{fig:korista}
\end{figure}

The specifics as to which particular radii contribute the most to the emission strength of each line is complicated to determine, and requires a more-through treatment of the ionization state of the gas as a function of radius, which depends on the geometry of the structure, and the distribution of density and temperature as functions of height and radius. In a tidal disruption, the matter distribution that ensheathes the black hole is established quickly, forming a steady-state structure that is supported by a combination of gas pressure and angular momentum \citep{Loeb:1997jv}. Accretion then proceeds through the midplane, in which the majority of light is generated within a few $r_{\rm g}$ at X-ray temperatures. These photons are intercepted by the ensheathing material at higher latitudes. \citet{Korista:2004cp} determined the equivalent widths of various lines as functions of volume density and ionizing flux, which is not expected to vary much as a function of column density for $10^{23} \leq N \leq 10^{25}$ cm$^{2}$ \citep{Ruff:2012tq}. In Figure \ref{fig:korista} we show a series of density-ionization curves corresponding to our highest-likelihood model for \psone over the range of times at which spectra were taken of this event. These are compared to the equivalent widths measured To calculate the density distribution $n({\rm H})$ as a function of $r$, we once again use the chain rule,
\begin{align}
n({\rm H}) &= \frac{X({\rm H})}{4\pi m_{\rm p} r^{2}} \frac{dM}{dE}\frac{dE}{da} \left(\frac{2 a}{r_{\rm p}}\right)^{3/2}\label{eq:nh}
\end{align}
where $X({\rm H})$ is the mass fraction of hydrogen. and presuming that the radial distribution of mass is determined by the distribution of mass with semi-major axis $a$, $dM/da$, set at the time of disruption. Likewise, $dM/dr$ is directly proportional to $dM/da$, with a scaling factor equal to the ratio of time spent at apocenter versus pericenter, \smash{$dM/dr \simeq dM/da (2 a / r_{\rm p})^{3/2}$}, where we have assumed that $1 - e \rightarrow 0$ and thus the apocenter distance $r_{\rm a} \simeq 2 a$.

As a strong dissipation mechanism likely operates at the nozzle point, and this dissipation mechanism is likely to be as dissipative as the commonly invoked MRI mechanism, it stands to reason that the vertical structure of the debris disk formed through the circularization process is similar to that of a steadily-accreting AGN. Therefore, we would expect that the BLR associated with such structures should be similar to the BLR produced by steadily-accreting black holes. Under this assumption, we can use Equation \ref{eq:nh} to approximate the number density of hydrogen as a function of radius and to determine the equivalent width of various emission lines using the models that have been generated for steadily accreting AGN \citep{Korista:2004cp}. Figure \ref{fig:korista} shows the density-ionizations curves calculated from Equation \ref{eq:nh} as a function of time for our highest-likelihood model of \psone, with the purple curve corresponding to the time of the first acquired spectrum of \psone at -22 days, and the red curve corresponding to the last recorded spectrum at +358 days.

From Figure \ref{fig:korista}, it is clear that the equivalent width of \heii is significantly larger than that of \ha, \hb, and \hei, all three of which are not observed in \psone. The figure does suggest that hydrogen and/or singly-ionized helium emission lines may appear at later times when the ionizing flux has decreased, although this may not ever be observable in \psone where the flux originating from the TDE has already dropped below that of the host galaxy.

In generating this plot, we have made some assumptions that actually would lead to a {\it decrease} in the strength of the unobserved lines if we performed a more-detailed calculation. Firstly, the models of \citet{Korista:2004cp} presume that a full annulus of locally optimally emitting clouds exists at each radius; this is not the case in an elliptical accretion disks where the inner annuli are closer to full circles than outer annuli \citep{Eracleous:1995fv}. In fact, it is unlikely that the outer material can circularize at all, given that there is significantly less angular momentum in the disk than the angular momentum required to support a circular orbit at the distance at which these lines would be produced (at $r = 10^{16}$ cm, $\Sim 30$ times more angular momentum would be required to form a circular orbit than what is available at $r_{\rm p}$). Secondly, we have made the assumption that the material that does the reprocessing remains at the distance determined by the energy distribution set at the time of disruption {\it at all times} \citep[\`{a} la][]{Loeb:1997jv}, when in reality the entire debris structure will shrink onto the black hole due to dissipation at pericenter. It is possible that this shrinkage of the debris could prevent emission features arising from species with lower ionization potentials from ever being observed.

Radiation pressure (which we ignore in this work) may act to push some fraction of the material outwards, which in principle could produce low-energy emission features \citep{Strubbe:2009ek}. However, our highest-likelihood models predict a peak accretion rate that is sub-Eddington, and thus only a small fraction of the accreted matter is expected to be driven to large distances via radiation. It is unclear whether the amount of mass in this component would be dense enough to produce these features, as the recombination time may be too long.

\begin{figure*}
\centering\includegraphics[width=0.54\linewidth,clip=true]{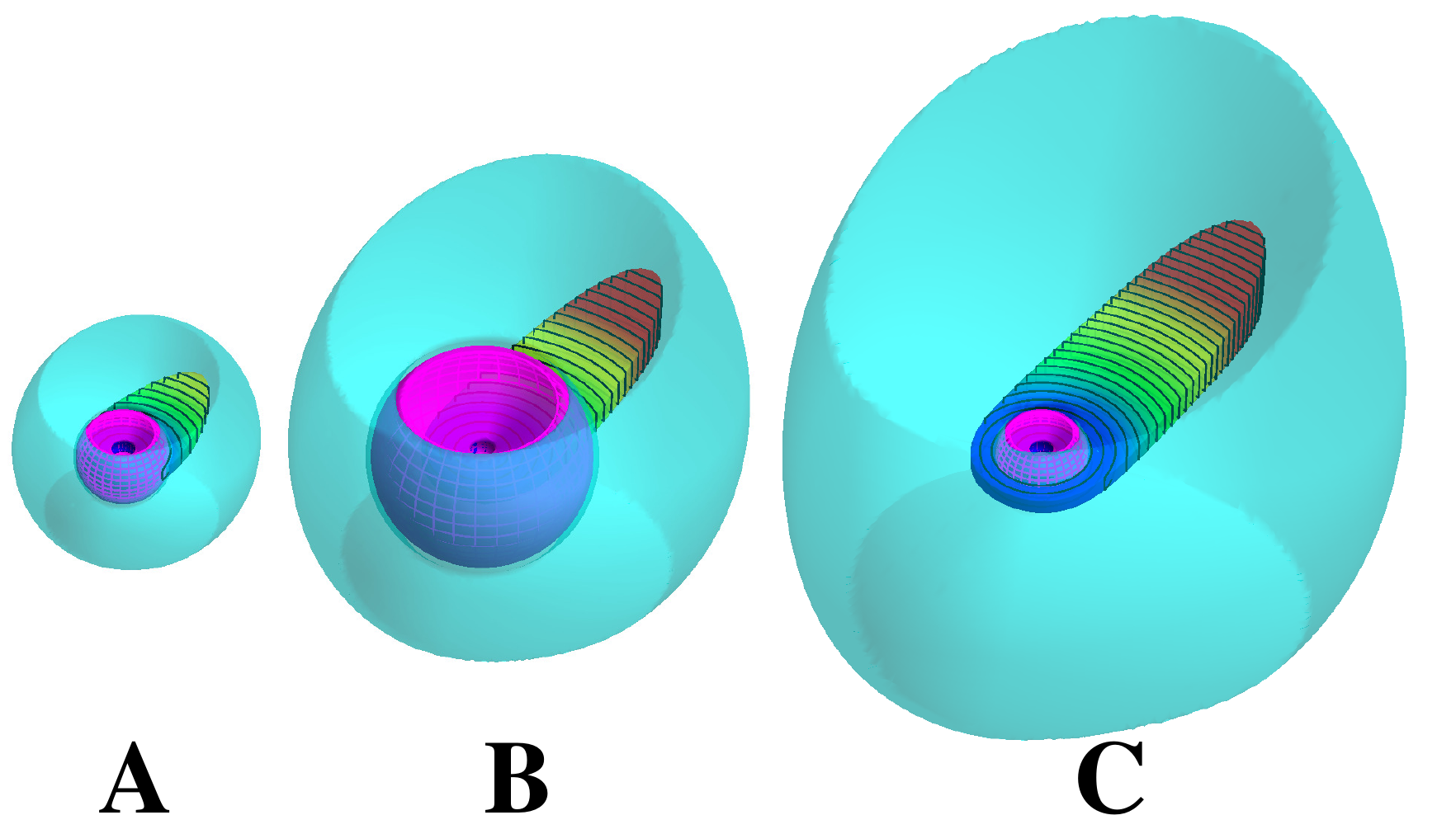}\includegraphics[width=0.46\linewidth,clip=true]{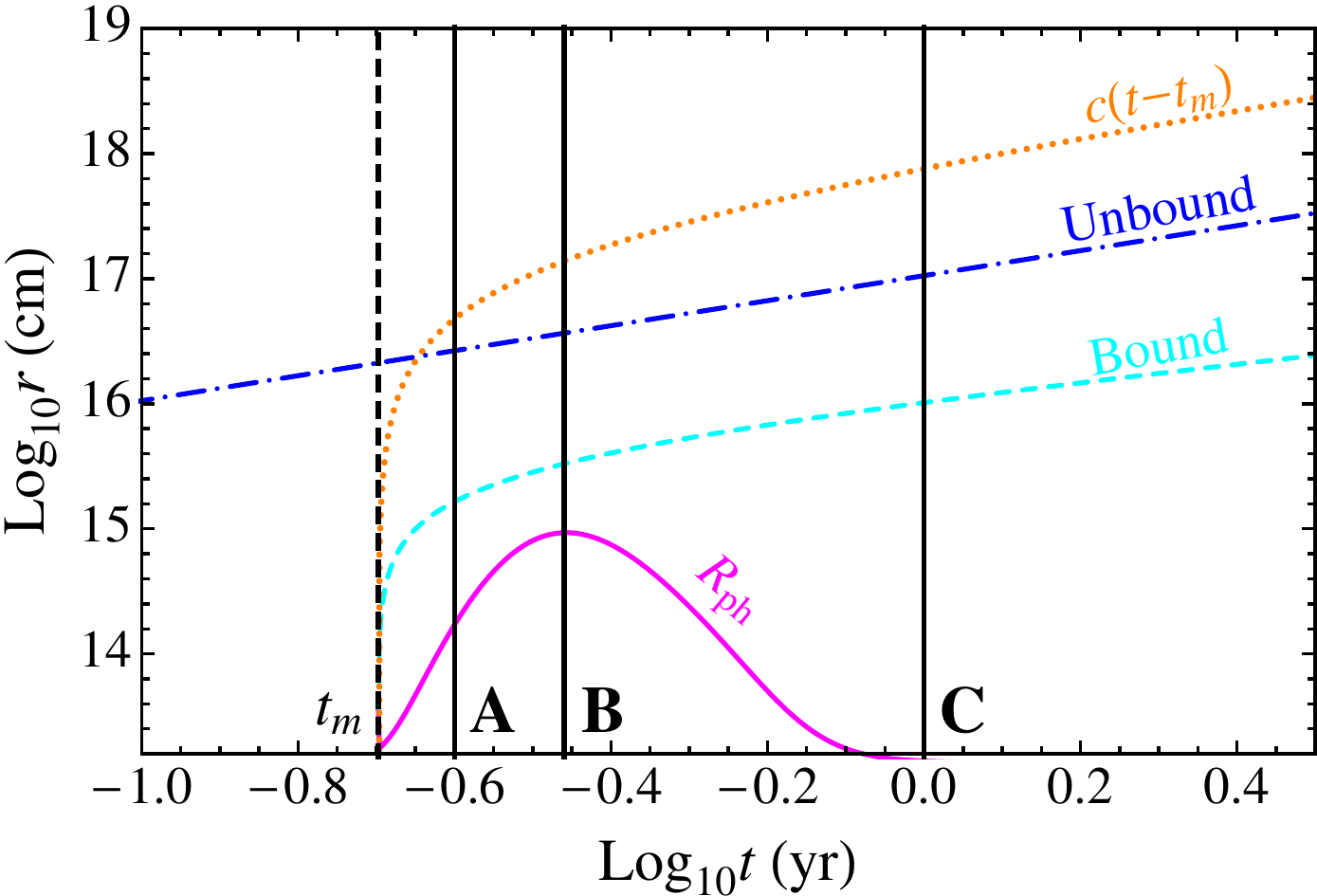}
\caption{Evolution of the size scales relevant to the appearance of TDEs. The left image shows a three-dimensional schematic of the elliptical accretion disk (rainbow-colored surface), the low-density material that ensheaths the disk (cyan surface), and the location of the region interior to which helium is doubly-ionized (magenta surface), at three times labeled A, B, and C. The right plot shows results from the highest-likelihood fit of \psone, where the solid cyan and dashed magenta curves correspond to the two surfaces in the left image, the dash-dotted blue curve corresponds to the distance to which the unbound material has traveled, and the dotted orange curve corresponds to the distance to which light travels since the time of the accretion of the most-bound material, denoted by the vertical dashed black line. The times to which the images on the left correspond are shown by the labeled vertical black lines.}
\label{fig:threed}
\end{figure*}

\section{A Generalized Model for the Observational Signatures of TDEs}\label{sec:model}

As emphasized in the previous sections, there are many uncertainties relating to how the material circularized when it returns to pericenter, how this returning material radiates its energy when it falls deeper into the black hole's potential well, and in the ionization state of the gas within the debris superstructure. Using the code \tdefit, developed for this paper, we construct a generalized model of the resultant emission from TDEs. In this section, we describe the results of running this fitting procedure, and how \psone specifically allows us to evaluate some of the other models that have been proposed for modeling TDEs.

\subsection{Model description and free parameters}\label{subsec:modeldesc}

Our generalized model for matching TDEs is one in which an accretion disk forms by the disruption of a star of mass $M_{\ast}$ by a black hole of mass $M_{\rm h}$ with impact parameter $\beta$ and offset time $t_{\rm off} \equiv t_{0} - t_{\rm d}$, where $t_{0}$ is the time \psone was first detected in Pg (May 10.55, 2010). This disk spreads both inwards and outwards from $r_{\rm p}$, and is ensheathed by a diffuse layer of material that intercepts some fraction of the light. The disk itself is bounded by an inner radius $r_{\rm i}$ and outer radius $r_{\rm o}$, with $r_{\rm i}$ assumed to be set by the viscous evolution of the material, and $r_{\rm o}$ being set by the ballistic ejection of material as it leaves the nozzle region, which scales as \smash{$r_{\rm o} = r_{\rm p} (t/t_{\rm m})^{2/3}$}, where $t_{\rm m}$ is the time of return of the most-bound material. The fraction of the full annulus $\theta_{\rm f}$ that is covered by the disk varies as a function of time, with $\theta_{\rm f} = 0$ when $r = r_{\rm o}$, and $= 2 \pi$ when $t = t_{\rm visc} \left(r\right)$, assuming its spread in the azimuthal direction is controlled by the local value of the viscosity. The model is shown pictographically in the left panel of Figure \ref{fig:threed}, with the aforementioned size scales as functions of time being shown in the right panel of the same figure.

The source of this viscosity may be similar to the source of viscosity at the nozzle point (see Section \ref{subsec:dissipative}), or it could be the result of the stream-stream collision that occurs when material reaches apocenter \citep{Kochanek:1994bn,Kim:1999dw,RamirezRuiz:2009gw}. For simplicity, we assume that the same viscous process, parameterized by the free parameter $\V$, applies in both regions. We presume that $\V$ is time-independent, resulting in a simple time-shift of $\md_{\rm acc}$ relative to $\md_{\rm fb}$, $\md_{\rm acc}(t/\V) = \md_{\rm fb}$, where $\md_{\rm acc}$ is the accretion rate onto the blackhole and is normalized such that the total mass accreted is equal to the integral over $\md_{\rm fb}$, the input fallback rate. For $\V = 1$, $\md_{\rm acc} = \md_{\rm fb}$, i.e. $t_{\rm visc} \lesssim t_{\rm m}$.

The emergent emission from the disk is calculated using the prescription of \citet{Done:2012eq}, which largely follows the original prescription of \citet{Shakura:1973uy}, but amends the no-torque boundary condition to include the effects of the black hole's spin, parameterized by the dimensionless spin parameter $a_{\rm spin}$. However, it is not immediately clear that the elliptical disk component, which is in the process of circularizing and spans from $2 r_{\rm t}$ to $r_{\rm o}$, would be adequately described by such models. For tidal disruption disks, the densities are low enough such that radiation pressure dominates, but high enough to be optically thick. If we presume that all material returns to pericenter cold, but then is heated to some degree by the circularization process at $r_{\rm p}$, the specific internal energy of fluid  at pericenter $\epsilon = (1/2) \rho \alpha v_{\rm K}^2$, which yields a temperature $T = [\alpha v_{\rm K}^2/(\rho a_{\rm rad})]^{1/4}$, where $a_{\rm rad}$ is the radiation constant and $\rho$ is the local density. As the scale height $H \propto \alpha$ near Eddington \citep{Strubbe:2009ek} and $\rho \propto 1/(r^{2} H)$, $T \propto (v_{\rm K}^2/a_{\rm rad} r^{-3})^{1/4}$. This is proportional (modulo a constant) to the temperature of a steadily-accreting thin disk at the same $r$ \citep{Beloborodov:1999wb}. After this injection of energy, the internal pressure of this fluid is balanced at all radii with tidal pressure along its elliptical trajectory, which scales as $r^{-3}$ when $H/R$ is fixed, and thus the temperature scaling with radius is identical to that of a steadily-accreting disk. As the inner circular accretion disk necessarily exchanges energy with the outer elliptical disk at their interface at $r_{\rm p}$, the temperature at this interface is likely to equilibrate to the inner disk's temperature at $r_{\rm p}$; we assume that this sets the normalization constant of the elliptical disk equal to the inner disk. Under this assumption, the temperature structure within the disk would follow \citet{Done:2012eq} verbatim.

The model of \citet{Done:2012eq} accounts for a shift in the emergent disk spectrum arising from variations in opacity, resulting in an effective temperature that can be $\Sim 2.7$ times larger than expected from the fiducial SS model. We do not permit the luminosity $L$ to exceed the Eddington luminosity $L_{\rm Edd} = 4 \pi G M_{\rm h} m_{\rm p} c / \kappa_{\rm t}$, where $\kappa_{\rm t}$ is the Thomson opacity $\kappa_{\rm t} = 0.2 (1 + X(H))$ cm$^{2}$~g$^{-1}$, and set $L = L_{\rm Edd}$ at times where $\md$ exceeds this limit. We also include the inclination of the structure relative to the observer $\phi$ as a free parameter, where $\phi = 0$ is defined to be edge-on, assuming that both the disk's height and the ensheathing layer scale with $\V$ in the same way, with the emergent emission from both components being reduced by a factor $\V + (1 - \V)\cos \left(\phi\right)$.

Note that we assume the color correction is intrinsic to the disk emission, and is not the same as the reprocessing that occurs due to the diffuse gas that ensheaths the disk and is ejected from the nozzle region directly. The photosphere of this reprocessing layer, whose size is set by a combination of the mass distribution and the absorption process responsible for intercepting the light, is less constrained. For steadily-accreting AGN with thermal emission, the reprocessing layer has temperatures of several $10^{4}$ K \citep{Koratkar:1999di,Lawrence:2012hn}, and intercepts nearly 100\% of the emission from the disk, resulting in an effective photosphere size that can be hundreds of AU in size for SMBHs accreting at the Eddington limit. However, as there are many non-thermal AGN whose spectra are more representative of bare slim-disk models \citep{Walton:2013ds}, it remains unclear how the size of this reprocessing zone and its fractional coverage are set.

In the models of \citet{Strubbe:2009ek,Strubbe:2011iw}, this layer is presumed to arise as the result of ejection via a super-Eddington wind, and scales with this value when $\md$ exceeds $\medd$. From our hydrodynamical simulations, we find that the material that forms the reprocessing layer may be deposited by a process that does not require the accretion rate to exceed $\medd$, but instead depends on the details of how energy is injected into the material within the nozzle region. The distribution of mass in radius resulting from the ejection from the nozzle maps is directly related to $\dmde$, although it is modified somewhat by the additional spread in energy introduced at the nozzle point. However, this spread in energy is local to mass that return at a particular time $t$, and thus the distribution of mass with radius after leaving the nozzle point will resemble $\dmde$ with an additional ``smear'' equal to the spread in energy applied at the nozzle. Therefore, we expect that the mass distribution with radius follows the general shape of $\dmde$, and that there will be a density maximum corresponding to the orbital period of the material that constitutes the peak of the accretion. This peak in density that corresponds to the apocenter of the material that determines \smash{$\mdpeak$} is clearly seen in our hydrodynamical simulations (Figure \ref{fig:columndensity}).

Thus, the size of the reprocessing layer is likely to be dependent on both the instantaneous value of $\md$, which determines the amount of ionization radiation produced by the disk and the rate of instantaneous mass loss from the nozzle region, and on the integrated amount of mass that has been ejected from the nozzle region since $t = t_{\rm d}$. The optical depth $\tau$ is
\begin{equation}
\tau = \kappa \int_{0}^{\infty} \frac{dM}{dr} dr\label{eq:tau},
\end{equation}
where $\kappa$ is an opacity that is at minimum $\kappa_{\rm t}$.

As we described in the derivation of Equation \ref{eq:nh}, the amount of mass at a particular distance $r$ is related to the amount of mass at a particular binding energy $E$, and thus we can rewrite Equation \ref{eq:tau} in terms of $E$,
\begin{equation}
\tau = \kappa \int_{E_{\rm m}}^{E_{\rm o}} \frac{dM}{dE} dE\label{eq:tau2}.
\end{equation}
where $E_{\rm m}$ and $E_{\rm o}$ are the binding energy of the most bound material and the material at apocenter at time $t$ respectively.

For simplicity, we presume that the reprocessing layer intercepts a fixed fraction of the disk's light, with the fraction of light $C$ intercepted by the reprocessing layer simply scaling with the optical depth $\tau$,
\begin{equation}
C = 1 - e^{-\tau},
\end{equation}
where $\tau$ is a free parameter. We enforce the condition that $C < c(t - t_{\rm d})/R_{\rm ph}$ (where $R_{\rm ph}$ is the size of the photosphere) at all times, otherwise the photon diffusion time would be greater than the time since disruption, and thus $L$ and $\md$ would not be expected to closely trace one another. In reality, $C$ should have a wavelength dependence, but for the purposes of this work we treat the opacity as being ``gray,'' absorbing all frequencies of light equally.

As the ionization state of the gas (and therefore the opacity) depends on the current luminosity, the size of the photosphere is expected to vary with time. In general, as the Thomson cross-section is significantly smaller than that of bound-free transitions, the photosphere scale is likely to correspond to the first species is not completely ionized, and in the case of \psone where \Heii emission is observed, we speculate that this species is helium (Figure \ref{fig:threed}, magenta region). If we assume that $R_{\rm ph} \propto \md^{l}$, $T_{\rm ph} \propto \md^{m}$, and $L \propto \md \propto R_{\rm ph}^{2} T_{\rm ph}^{4}$, then the power law indices of $R_{\rm ph}$ and $T_{\rm ph}$ are simply related,
\begin{equation}
2l + 4m = 1.\label{eq:powlaws}
\end{equation}
If the opening angle of the reprocessing layer is independent of $r$, the flux in ionizing photons intercepted is constant, implying $l = 1/2$ and thus $m = 0$, i.e. $T_{\rm ph}$ is independent of time. However, as we find that the geometry may in reality be somewhat more complicated (Figure \ref{fig:columndensity}), we do not assume the intercepting area necessarily scales as $\md$, and instead leave $l$ as a free parameter. For any $l \neq 1/2$, the temperature of the photosphere will evolve with time. We leave $l$ as a free parameter and relate $m$ and $l$ through Equation \ref{eq:powlaws}. The size of the photosphere is then defined to be
\begin{align}
R_{\rm ph} &= R_{\rm ph,0} a_{\rm p} \left(\frac{\md}{\medd}\right)^{l}\label{eq:rph}\\
a_{\rm p} &= \left[8 G M_{\rm h}\left(\frac{t_{\rm peak} - t_{\rm m}}{\pi}\right)^{2}\right]^{1/3},
\end{align}
where $a_{\rm p}$ is the semi-major axis of the material that accretes at $t = t_{\rm peak}$, and $R_{\rm ph,0}$ is a dimensionless free parameter.

The amount of reddening in the host galaxy is also an unknown quantity that must be fitted to simultaneously with the parameters of the disruption. For extinction in the IR through the UV, we adopt the reddening law fits of \citet{Cardelli:1989dp}, in which the amount of reddening is defined by $A(\lambda) = A_{\rm V} [a(\lambda) + b(\lambda)/R_{\rm V}]$, in which $a(\lambda)$ and $b(\lambda)$ are fitted parameters, $R_{\rm V}$ is a fitted parameter that ranges between 2 and 10 \citep{Goobar:2002ea}, and where we take $N_{\rm h} = 1.8 \times 10^{21} A_{\rm V}$ g~cm$^{-3}$. For the X-rays, we adopt the cross-sections presented in \citet{Morrison:1983bh}. Extinction in the X-rays is particularly sensitive to metallicity and temperature \citep{Gnat:2012gz}, and the uncertainty in the amount expected for a particular event is large given the environment of a galactic center is likely to have super-solar metallicities \citep{Cunha:2007gp} and a wide range of temperatures and densities \citep{Quataert:2002jc,Cuadra:2006ju,DeColle:2012bq}.

An advantage of the MCMC method employed here is that it permits the inclusion of discrete parameters that can only assume particular values. This enables us to simultaneously fit multiple physical models, as long as the continuous parameters are shared between the models. We include two discrete free parameters in this work: ${\cal A}_{\ast}$, which parameterizes the type of object that was disrupted, and ${\cal A}_{\gamma}$, which parameterizes the polytropic model that is assumed. We include two distinct object types, the white dwarf sequence and the main sequence. Within each of these sequences, different mass ranges are characterized by different polytropic $\gamma$; we use the $\md$ functions derived from our hydrodynamical simulations \citepalias{Guillochon:2013jj} appropriate to each mass range,
\begin{align}
\md =\; &\dot{M}_{4/3}\left(t\right) \left\{
\begin{array}{ll}
{\rm MS}&:
	\begin{array}{ll}
	&0.3 < M_{\ast}/M_{\odot} < 22\\
	\end{array}\\
{\rm WD}&:
	\begin{array}{ll}
	&M_{\ast}/M_{\odot} > 1.0
	\end{array}\\
\end{array}
\right.\\
\md =\; &\dot{M}_{5/3}\left(t\right) \left\{
\begin{array}{ll}
{\rm MS}&:
	\begin{array}{ll}
	&M_{\ast}/M_{\odot} < 1.0\\
	&M_{\ast}/M_{\odot} > 22
	\end{array}\\
{\rm WD}&:
	\begin{array}{ll}
	&M_{\ast}/M_{\odot} < 1.0
	\end{array}\\
\end{array}
\right.
\end{align} 
where some overlap is permitted in the mass range $0.3 < M_{\ast}/M_{\odot} < 1.0$ to account for the gradual transition between fully radiative and fully convective stars in this range. It was found that the white dwarf sequence, which is only permits very low mass black holes, is excluded to very high confidence for all of the combinations of parameters that were considered, especially when accounting for the measurement of the black hole's mass presented in \citetalias{Gezari:2012fk}, which restricts $M_{\rm} > 2 \times 10^{6} M_{\odot}$. For simplicity, we exclude discussion of the white dwarf channel for the rest of this work.

\begin{figure*}
\centering\includegraphics[width=0.9\linewidth,clip=true]{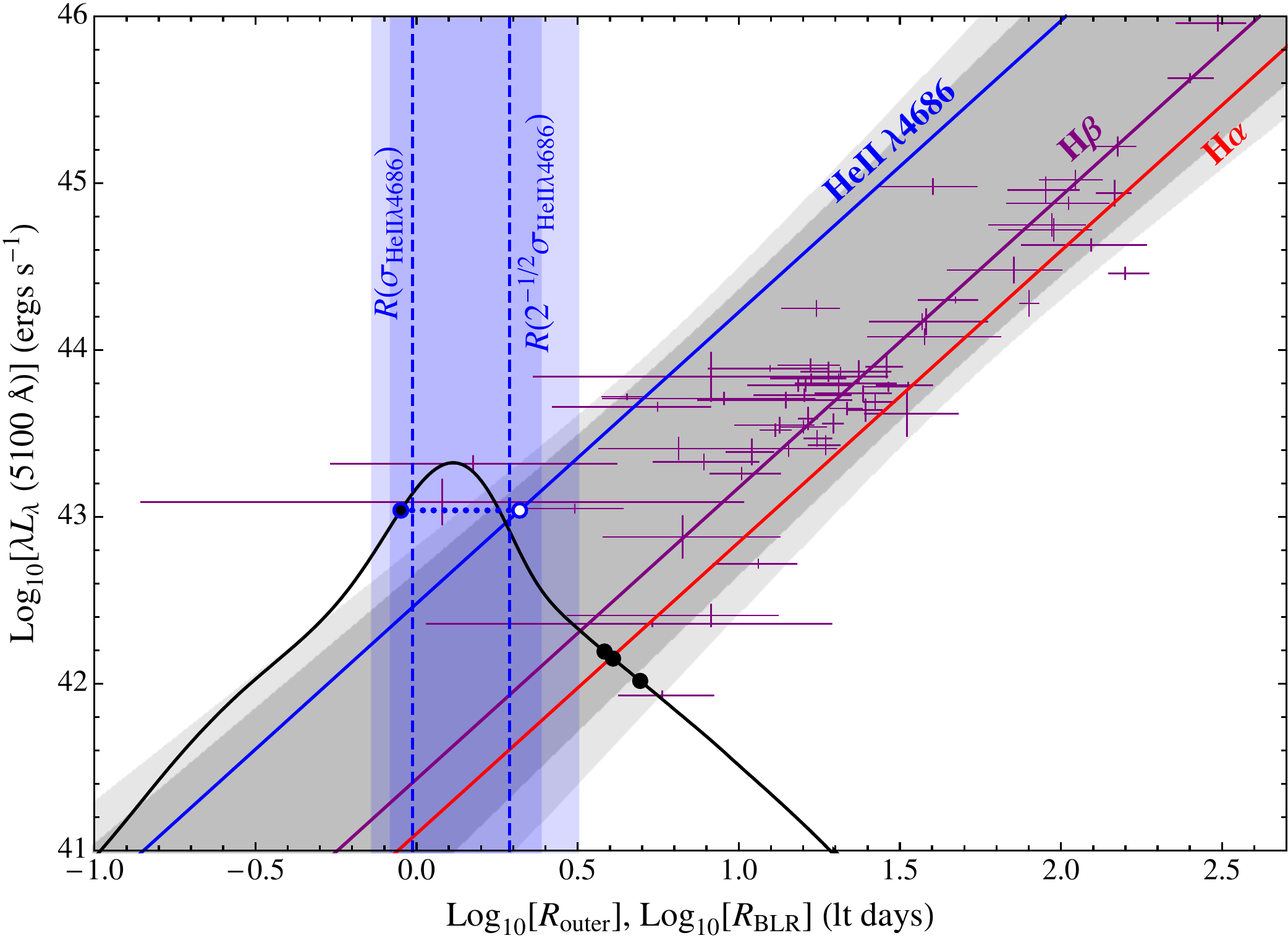}
\caption{Location of BLR in AGN as a function of $L$ at $\lambda = 5100$ \AA\, as compared to the outer radius of the truncated debris disk resulting from a tidal disruption. The purple error bars and purple curve show the data and fit of \citet{Peterson:2006hr} for \hb, while the red and blue curves show the offsets to the best-fitting H$\beta$ relation as measured by \citet{Bentz:2010hk} for \ha and \heii. The outer radius of the truncated debris disk $r_{\rm o}$ is shown with the black curve, with the time of collection for the four spectra (-22, +227, +254, and +358 restframe days) being denoted by the black points. In all but the last spectrum, in which no lines are apparent, \heii was observed and \ha was not observed, which implies that $r_{\rm o}$ must extend beyond the region within which the majority of the flux for \heii is produced, but not extend beyond the region within which most of the \ha flux is produced. Thus, the four black points must appear between the red and the blue curves to satisfy this constraint (shaded in gray). Additionally, the velocity dispersion of the \heii line ($\sigma_{\rm HeII}$) was measured for the first spectrum taken at -22 days. The \heii line is assumed to be located along the standard BLR relation, and the velocity of this line predicted by the model is given by matching the luminosity to its associated size on the \heii curve (white point). As it is unclear whether the motions being observed are Keplerian or inflow/outflow, the distance implied by the model velocity is constrained between a circular and a parabolic orbit, \smash{$G M_{\rm h} / \sigma_{\rm HeII}^{2} < R_{\rm HeII} < 2 G M_{\rm h} / \sigma_{\rm HeII}^{2}$} (dashed vertical blue curves, shaded in blue).}
\label{fig:peterson}
\end{figure*}

In addition to modifying the emergent disk spectrum, $a_{\rm spin}$ also affects the minimum approach distance of a star on a parabolic trajectory (i.e. the innermost bound circular orbit) $r_{\rm IBCO}$, and the spread in energy across the star at pericenter \citep{Kesden:2012kv}. For simplicity in this work we only consider prograde encounters ($a_{\rm spin} > 0$), and apply first-order correction factor to the binding energy,
\begin{equation}
E^{\prime} = \left(1 - \frac{1}{2} \frac{r_{\rm IBCO}}{r_{\rm p}}\right)^{-1/2} E.
\end{equation}
In general, retrograde and/or orbits in which the orbit's inclination is not equal to the black hole's spin inclination would be expected.

AGN show variability from the radio to the X-ray, with variability on the order of a few tenths of a dex being common \citep{Webb:2000iv}. While the photometric errors are small for this event, it is clear that the light curve exhibits some intrinsic variability, as may be expected for an accreting black hole. To model this, we add an additional intrinsic spread $\sigma_{\rm v}$ in quadrature with the observational errors associated with each data point. In addition, the variability has been shown to be dependent on the black hole mass \citep{Uttley:2005jq}.

In total, our fitting procedure includes 15 parameters, 13 of which are continuous ($M_{\ast}$, $M_{\rm h}$, $\beta$, $t_{\rm off}$, $a_{\rm spin}$, $\V$, $\phi$, $\tau$, $l$, $R_{\rm ph,0}$, $R_{\rm v}$, $N_{\rm h}$, and $\sigma_{\rm v}$), and 2 of which are discrete (${\cal A}_{\gamma}$ and ${\cal A}_{\ast}$).

\subsection{Using the existence/absence of emission lines and their properties to constrain TDEs}
In addition to using the quality of the fit of the model light curves to the data, we also impose additional constraints depending on which lines do or do not exist in spectra taken at various times (Figure \ref{fig:peterson}). To do this, we measure the emergent flux at $\lambda = 5100 \AA$, which is used in steadily-accreting AGN to measure the continuum, and compare $r_{\rm o}$ to the distance implied by the relationship between $\lambda L_{\lambda} (5100 \AA)$ and $R_{\rm BLR}$, as first determined for \hb by \citet{Wandel:1999ev}. Since then, the relationship between $L$ and $R_{\rm BLR}$ has been more-accurately determined for \hb \citep{Peterson:2004ig,Bentz:2013hm}, and for several other emission lines \citep{Bentz:2010hk}. For all lines, it is found that $L \propto R_{\rm BLR}^{0.5}$, indicating that the source of ionizing photons is point-like and that the disk maintains a relatively constant scale-height for a wide range of $r$, which gives the natural result that the number of ionizing photons $\Phi \propto r^{-2}$.

Our modification to the likelihood function is simple: If a line exists in a spectrum and $r_{\rm o} < R_{\rm BLR}$, or if a line doesn't exist and $r_{\rm o} > R_{\rm BLR}$, we reduce the log-likelihood measured from the light curve alone ${\cal L}_{\rm LC}$ by a factor
\begin{equation}
\ln {\cal L_{\rm BLR}} = \ln \left[1 - \frac{1}{2}{\rm erfc}\left(\frac{\left|\ln R_{\rm BLR} - \ln r_{\rm o}\right|}{2 \sigma_{\rm BLR}}\right)\right],\label{eq:lblr}
\end{equation}
where $\sigma_{\rm BLR}$ is the error in the measured $L-R_{\rm BLR}$ relation, which we take from \citet{Bentz:2010hk}.

If a line exists, and a velocity for that line has been measured, we can use that additional information to constrain the event further by relating $R_{\rm BLR}$ to the underlying velocity expected at the position. An uncertainty exists in TDE debris disks in that the underlying velocity $v_{\rm BLR}$ can range from Keplerian ($v_{\rm BLR} = v_{\rm K}$) to parabolic ($v_{\rm BLR} = \sqrt{2} v_{\rm K}$), and therefore we cannot constrain the distance implied by $v_{\rm BLR}$ better than a factor of $\sqrt{2}$. Bearing this in mind, our reduction to the log-likelihood assumes the following functional form,
\begin{equation}
\ln {\cal L}_{v} =\; \left\{
\begin{array}{ll}
	v < v_{\rm BLR}&\frac{1}{2}\left(\frac{v - v_{\rm BLR}}{\sigma_{\rm v}}\right)^{2}\\
	v_{\rm BLR} < v < \sqrt{2}v_{\rm BLR}&0\\
	v > v_{\rm BLR}&\frac{1}{2}\left(\frac{v - \sqrt{2}v_{\rm BLR}}{\sigma_{\rm v}}\right)^{2}.
\end{array}
\right.
\end{equation}

We show the results of imposing these constraints for \psone in Figure \ref{fig:peterson}, where we plot the distance to which material has traveled $r_{\rm o}$ versus the luminosity at 5100 \AA. The specific constraints we have applied are that \heii must be produced, and \ha must not be produced, in the four spectra in which the observed light is not dominated by the host galaxy (at -22, +227, +254, and +358 restframe days). We find that  $r_{\rm o}$ is sufficiently large to produce \heii in all four of these spectra, and that \ha would potentially be observable at later times if the host galaxy did not dominate the observed light (the light is already subdominant to the host galaxy at +254 days). \hb, which is produced at smaller distances than \ha in steadily-accreting AGN, may potentially be observable in late-time spectra, but its wavelength \smash{(4861 \AA)} notably overlaps with the observed broad \heii feature, and is usually a factor of $\Sim 3$ than \ha in most AGN \citep{Osterbrock:2006ul}. We also might expect that \hg and/or \hei may appear at later times, as the distances at which these lines are produced are only slightly larger than the distance at which \heii is produced \citep{Bentz:2010hk}. As we had mentioned in Section \ref{subsec:blrsource}, we may also be overestimating the size of reprocessing region if it accretes onto the black hole quickly, which would tend to predict the existence of more emission features. Given these uncertainties, we do not impose a constraint on the non-existence of \hei, \hb, or \hg in this work; we note that their inclusion would likely restrict the size of the accretion disk further, which would tend to favor lower-mass black holes.

We note that the constraints we are imposing do not consider the specific luminosity of the lines versus the continuum (see Section \ref{subsec:blrsource}), which can strongly affect whether a line is identified within a collected spectrum. This means we also cannot consider in detail the effects of the elliptical accretion disk structure resulting from a tidal disruption on the strengths of the observed lines, which would preferentially reduce the strength of lines originating at large distances as the BLR does not occupy a full $2\pi$ in azimuth in the outskirts of the debris structure (Figure \ref{fig:columndensity}).

\section{Model Fitting of \psone, a Prototypical Tidal Disruption}\label{sec:ps1-10jh}

\subsection{Available Data}

For the fitting procedure, we use all of the available data to constrain the event, including four {\it Pan-STARRS} bands (Pg, Pr, Pi, Pz), the X-ray upper limits from the {\it Chandra} space-based X-ray telescope (cycle 12), and the spectra taken by the Hectospec instrument on the MMT telescope, all of which are taken from \citetalias{Gezari:2012fk}. As the data presented in \citetalias{Gezari:2012fk} is already corrected for extinction assuming $N_{\rm h} = 7.2 \times 10^{19}$ cm$^{-2}$, we remove this correction before using the data as an input, as we self-consistently determine the extinction in the model fitting process. We assume a redshift $z = 0.1696$ as is determined in \citetalias{Gezari:2012fk} from template fitting to the host galaxy.

\subsection{Bare Disk Models}\label{subsec:baredisk}

In steadily-accreting AGN disks, the majority of radiation produced by the disk is thought to be intercepted by intervening gas that reprocesses the original emission from the disk. In a TDE, this layer may take some time to form, or may not form at all, depending on the dissipative processes at work. In this case, the light produced as the result of a TDE would resemble a bare Shakura-Sunyaev \citep[SS,][]{Shakura:1973uy} or slim-disk \citep{Abramowicz:1988bz} model, with peak emission that extends well beyond the tens of eV that is characteristic of AGN spectra. For this model, we do not include the additional constraints imposed by the existence/absence of emission lines.

\begin{figure*}
\centering\includegraphics[width=0.333\linewidth,clip=true]{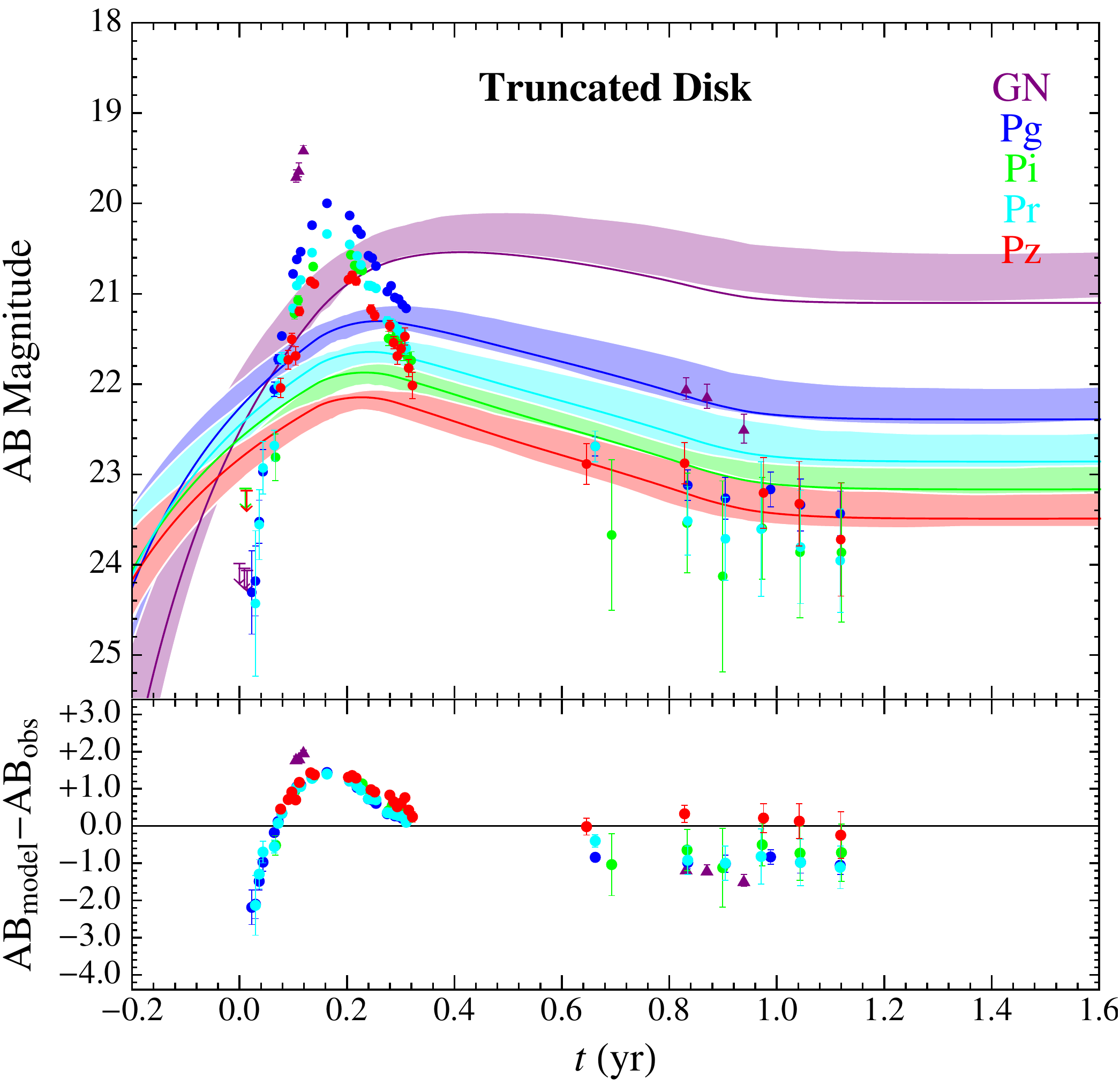}\includegraphics[width=0.333\linewidth,clip=true]{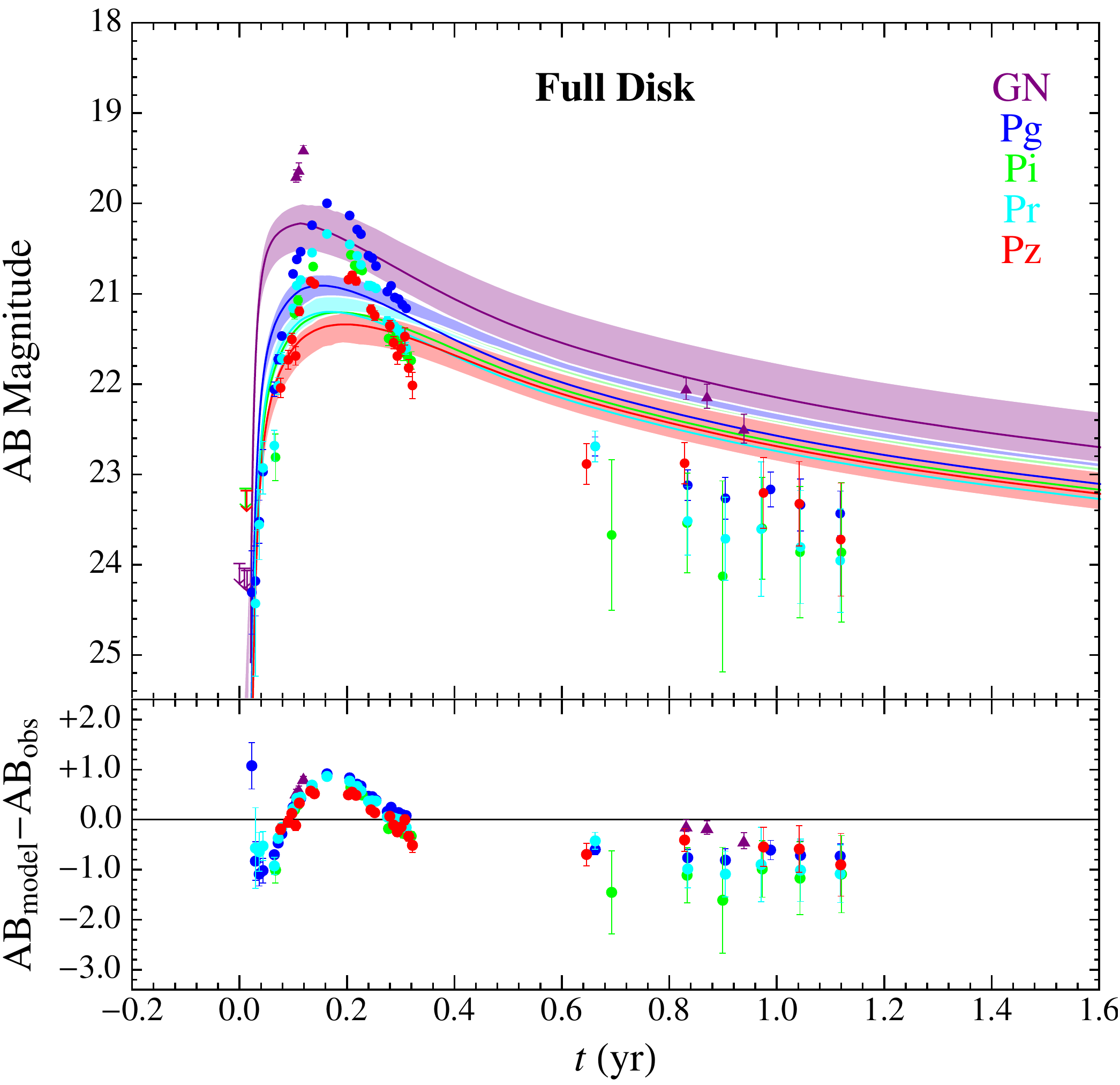}\includegraphics[width=0.333\linewidth,clip=true]{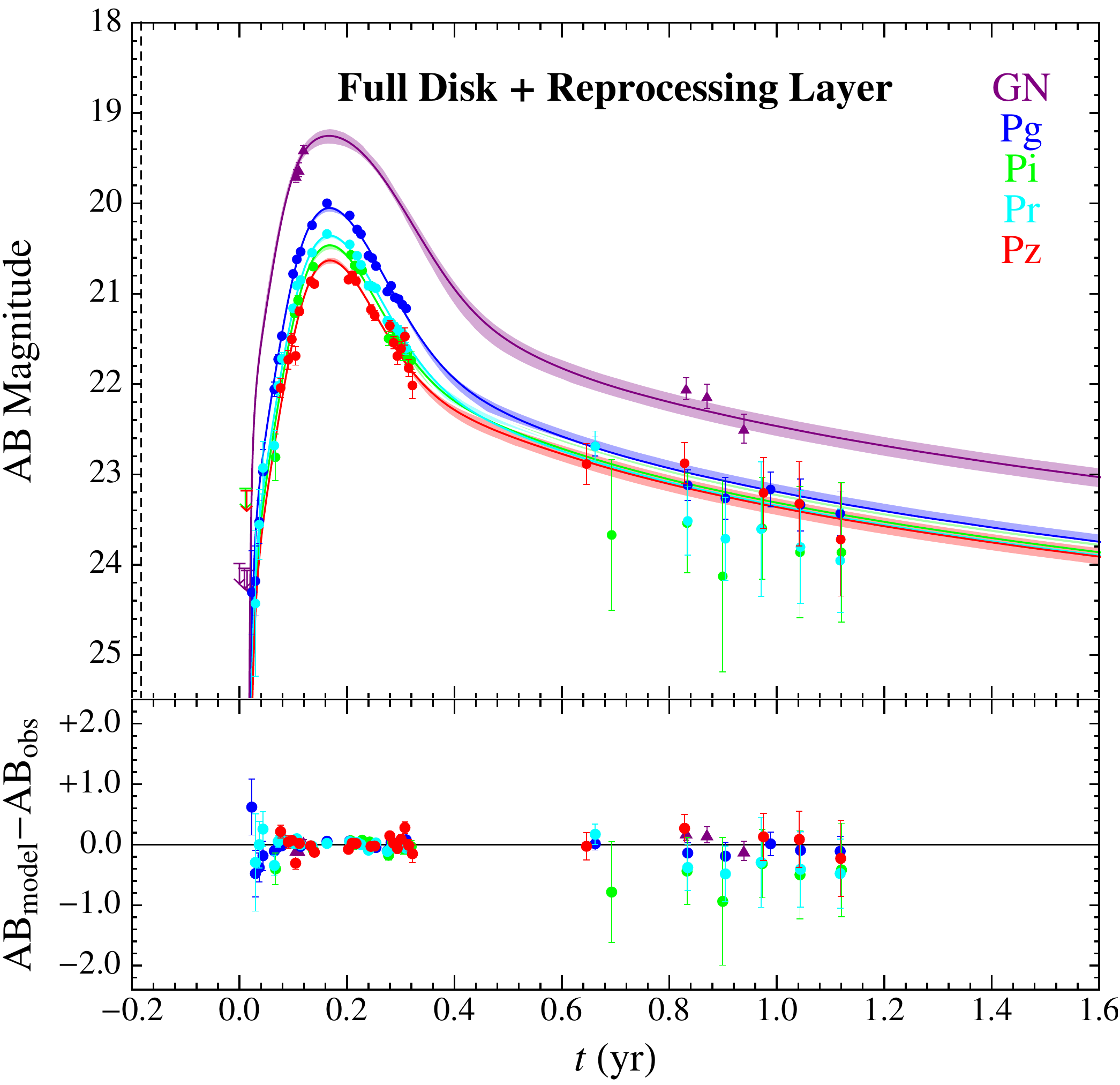}
\caption{Fits of highest likelihood of \psone for three different models: A bare SS disk truncated at $r = 2 r_{\rm p}$ (left panel, described in Section \ref{subsec:baredisk}), a bare SS disk that extends to $r_{\rm o}$ (middle panel, described in Section \ref{subsec:baredisk}), and a disk with a variably-sized reprocessing layer (right panel, described in Section \ref{subsec:generalmodel}), with the AB magnitude of the data and the models shown in the top panels, and the difference between the highest-likelihood model and the data shown in the bottom panels. The highest-likelihood model found is shown by the solid curves, whereas the 2-$\sigma$ range in magnitudes encompassed by the full ensemble of walkers is shown by the shaded bands. The five colors correspond to four filters of {\it Pan-STARRS} system (denoted by Pg, Pi, Pr, and Pz), and the NUV band of the {\it GALEX} instrument (denoted as GN).}
\label{fig:fits}
\end{figure*}

\begin{figure*}
\centering\includegraphics[width=0.5\linewidth,clip=true]{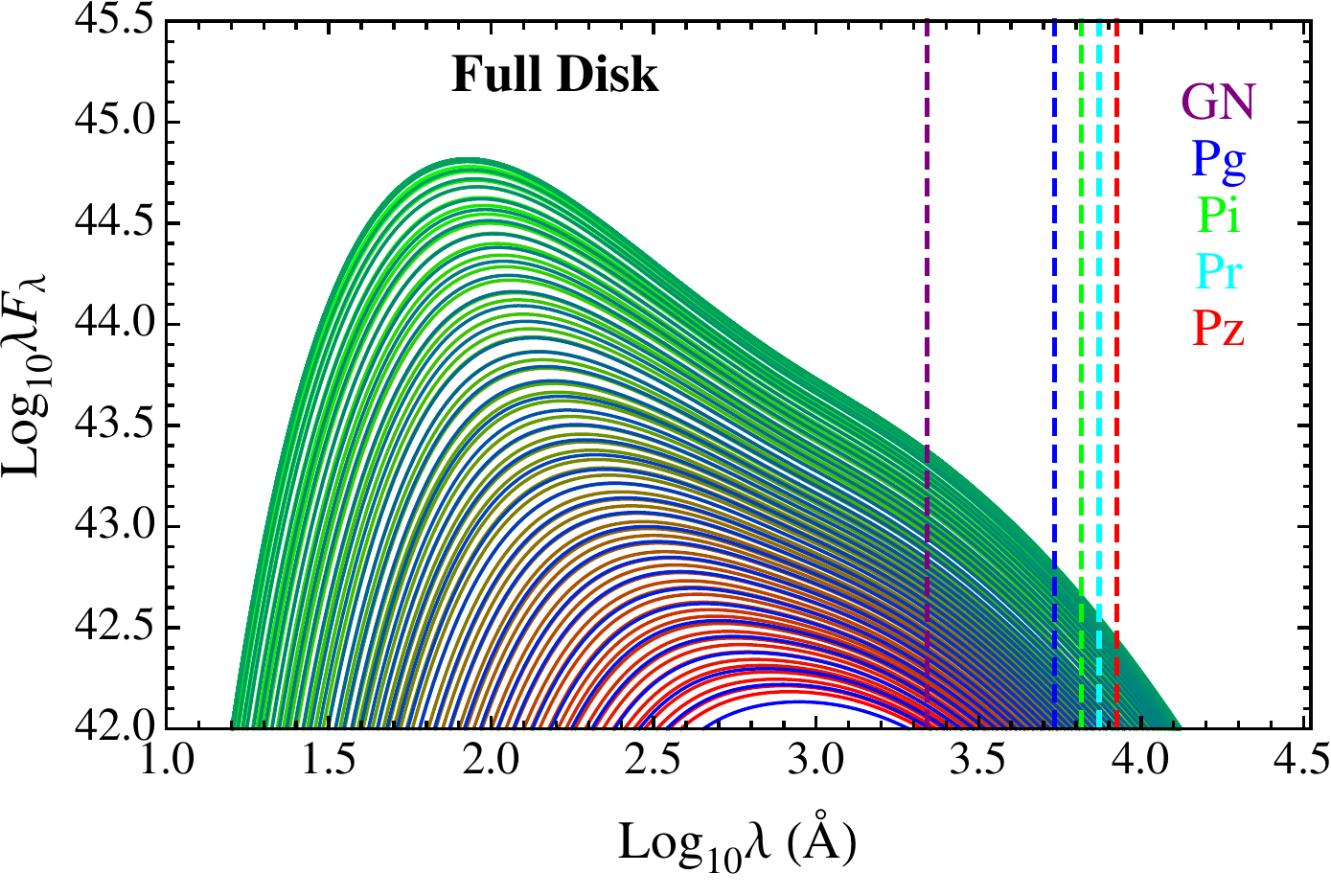}\includegraphics[width=0.5\linewidth,clip=true]{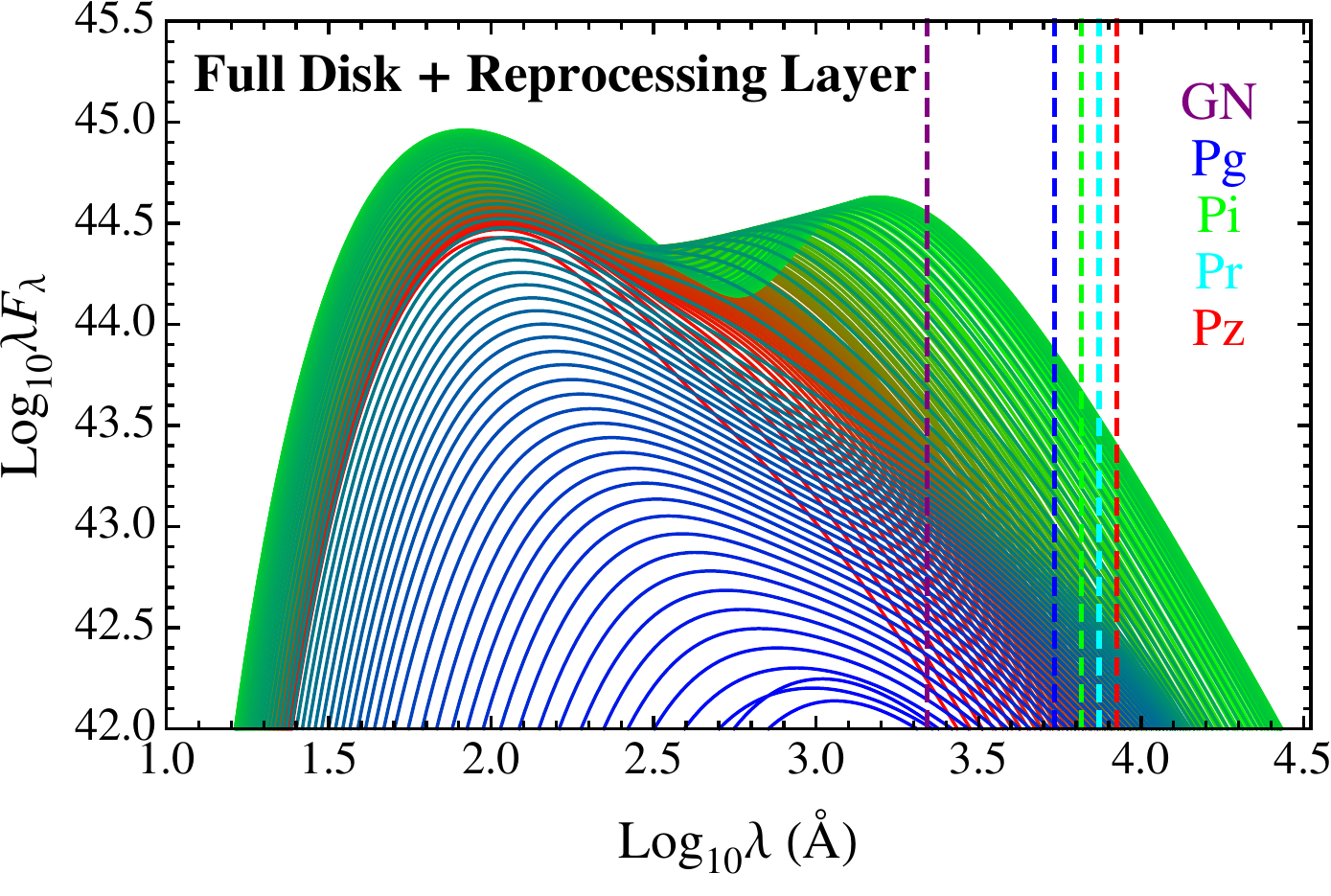}
\caption{Spectral energy distribution of the emergent radiation (before extinction from host galaxy or MW) from the highest-likelihood fits of \psone as a function of time for the bare disk (left panel) and generalized (right panel) models. The colored curves are spaced equally in $\ln\left(t - t_{\rm d}\right)$, with red corresponding to early times and blue correspond to late times. The vertical dashed lines show the wavelength of highest transmission for the four {\it Pan-STARRS} and {\it GALEX} NUV bands. The bare disk models feature a single component that corresponds to the accretion disk that peaks at \smash{$\lambda \sim 10^{2} $ \AA}, whereas the generalized model features two components which corresponds to the accretion disk, which peaks at \smash{$\lambda \sim 10^{2}$ \AA}, and the single-temperature reprocessing component, which peaks at \smash{$\lambda \sim 10^{3}$ \AA}.}
\label{fig:seds}
\end{figure*}

It has previously been assumed that the size of the disk is controlled by the angular momentum content of the returning material, which is limited to $\sqrt{2 G M_{\rm h} r_{\rm t}}$ (see Section \ref{subsec:stream} for references). In this case, the disk resembles a bare disk that is ``truncated'' at $r = 2r_{\rm t}$. We find that these models are a very poor match to the event (Figure \ref{fig:fits}, left panel).

Fits that do not truncate the disk and extend to $r_{\rm o}$, which are equivalent to our generalized model for TDE debris disks sans the reprocessing layer (cyan region of Figure \ref{fig:threed}), are shown in Figure \ref{fig:fits} (middle panel). This model is also a poor match to \psone, although the increase in surface area relative to the truncated disk does enable the model to at least reproduce \psone's peak luminosity. In order for the bare disk model to closely match the data, the fitting routine settles upon one of two non-ideal solutions: An SED with peak that centers about the range of wavelengths covered by the observed bands, or an SED in which the bands are all within the Rayleigh-jeans tail. In the first case, the luminosity $L$ can closely follow $\md$, but the color evolves tremendously as the peak of the summed blackbody curves shift into/out of the observed bands. In the latter case, the ratio of fluxes between the observed bands remains constant, but $L$ scales as a much weaker power of $\md$, $L \propto \md^{1/4}$ (Figure \ref{fig:seds}).

\subsection{Fits to Generalized Model With Reprocessing Layer}\label{subsec:generalmodel}

From the previous section, we know that bare disk models can either reproduce a constant color, or reproduce a luminosity that follows $\md$, but cannot reproduce both behaviors simultaneously. This suggests that a secondary process is involved that reprocesses a large fraction of the light prior to reaching the observer. In section \ref{subsec:generalmodel} we suggested that this mechanism is the absorption of the soft X-ray photons produced primarily at $r \sim r_{\rm g}$ by material deposited at $r \sim a_{\rm p}$.

As can be seen in the right panels of Figures \ref{fig:fits} and \ref{fig:seds}, these models provide excellent fits to the data; the parameters associated with the fits of highest likelihood are shown in Table \ref{tab:parameters}. We immediately caution the reader that the reported medians of the probability distributions, and the small spread in distributions of some parameters, should not be taken at face value. In our generalized model, which is only a simplified realization of the true structure of the debris, we have made many assumptions, and the uncertainty in some of these assumptions is likely to be greater than spread of solutions about the highest-likelihood models presented here. That being said, it is encouraging that such a simple model with relatively few free parameters can provide a reasonable fit to the data, and is highly suggestive of the true values of the underlying parameters. Note that the dominance of the reprocessing region over the emission from the disk in the optical/UV is similar to the model of \citet{Armijo:2013bo} in which an average temperature is calculated from the disk and used to fit \psone as a single, time-evolving blackbody.

\begin{deluxetable}{cccll} 
\tablecolumns{5}
\tablewidth{0pc}
\tablecaption{Parameters of Highest Likelihood Models}
\tablehead{
\colhead{Parameter(s)} & \colhead{Units} & \colhead{Prior} & \colhead{Allowed Range} & \colhead{Value\tablenotemark{a}}}
\startdata
$\Log_{10} M_{\ast}$ 		& $M_{\odot}$ 		& Flat & 
\phd\phn\phn$-3 \rsep 2$ & \phn\phn$0.576_{-0.143}^{+0.151}$ \\[2 pt]
$\Log_{10} M_{\rm h}$ 		& $M_{\odot}$ 		& Flat & 
\phs\phd\phn\phn$4 \rsep 8.6$ & \phs\phn\phn$7.25_{-0.08}^{+0.09}$ \\[2 pt]
$\beta$ 						& \nodata 			& Flat & 
\phs\phn$0.5 \rsep 4$ & \phs\phn\phn$1.32_{-0.02}^{+0.02}$ \\[2 pt]
$t_{\rm off}$ 				& days 				& Flat & 
\phd$-700 \rsep 700$ & \phs\phd\phn\phn\phn$77_{-8}^{+9}$ \\[2 pt]
$a_{\rm spin}$ 				& \nodata 			& Flat & 
\phs\phn\phn\phd$0 \rsep 1$ & \phs\phn\phn$0.37_{-0.26}^{+0.34}$ \\[2 pt]
$\Log_{10} \V$ 				& \nodata 			& Flat & 
\phd\phn\phn$-4 \rsep 0$ & $-0.18_{-0.05}^{+0.05}$ \\[2 pt]
$\phi$ 						& radians 			& Flat & 
\phs\phd\phn\phn$0 \rsep \pi/2$ & \phs\phn\phn$0.40_{-0.29}^{+0.36}$ \\[2 pt]
$\Log_{10} \tau$ 			& \nodata			& Flat & 
\phn$-6 \rsep 6$ & \phn\phn$-0.31_{-0.28}^{+0.27}$ \\[2 pt]
$l$ 							& \nodata 			& Flat & 
\phd\phs\phn\phn$0 \rsep 4$ & \phs\phn\phn$2.7_{-0.3}^{+0.3}$ \\[2 pt]
$\Log_{10} R_{\rm ph, 0}$ 	& \nodata 			& Flat & 
\phd\phn\phn$-4 \rsep 4$ & \phn\phn$0.17_{-0.19}^{+0.28}$ \\[2 pt]
$R_{\rm v}$ 				& \nodata 			& Flat & 
\phs\phd\phn\phn$2 \rsep 10$ & \phs\phn\phn\phn$6.5_{-0.4}^{+0.4}$ \\[2 pt]
$\Log_{10} N_{\rm h}$ 		& cm$^{-2}$ 			& Flat & 
\phd\phs\phn$17 \rsep 23$ & \phs\phn\phn\phn\phd$21_{-0.03}^{+0.03}$ \\[2 pt]
$\sigma_{\rm v}$ 			& \nodata 			& Flat & 
\phs\phd\phn\phn$0 \rsep 1$ & \phs\phn$0.05_{-0.007}^{+0.009}$
\enddata
\label{tab:parameters}
\tablenotetext{a}{Median value, with ranges corresponding to 1-$\sigma$ spread from median.}
\end{deluxetable}

In the generalized model without priors, we find that the disruption is best matched by the complete disruption ($\beta = 1.32$) of a moderate mass star ($M_{\ast} = 4 M_{\odot}$) by a $M_{\rm h} = 2 \times 10^{7} M_{\odot}$ black hole. This combination of parameters is close to the most common sub-Eddington disruption expected \citep{DeColle:2012bq}, but predicts the black hole mass is a factor of a few larger than the black hole mass suggested by \citetalias{Gezari:2012fk}. Most of this discrepancy is likely to arise not from improper template fitting of the host galaxy, but rather the large intrinsic scatter in the $M_{\rm h}$-$L$ relation, as black holes of mass $10^{6} \leq M_{\rm h}/M_{\odot} \leq 10^{9}$ have been found for other galaxies of similar magnitude \citep{Graham:2013fc}. The disruption is predicted to have occurred 42 days prior to the first observation, about 20 days prior to what was originally suggested in \citetalias{Gezari:2012fk}.

We find that $a_{\rm spin}$ is only loosely constrained, with the main effects of a larger spin being that deeper-$\beta$ encounters would be permitted (which are disfavored anyway), and an increase in the efficiency of converting mass to light. The inclination $\phi$ is highly degenerate with this parameter, and shows a strong anti-correlation (i.e. more-slowly spinning black holes tend to be more face-on). We find that the preferred models increase $\V$ to as large a value as possible, and likely this result would be altered given a physical model for $\V$ that accounts for all the various dissipation processes (see Section \ref{subsec:dissipative}).

If the mass from the disruption were spread evenly in azimuth, its optical depth would be quite large, $\tau \gtrsim \rho \kappa_{\rm t} r_{\rm p} \sim 100$, but the $\tau$ values returned by our fitting routine suggest that $\tau \simeq 0.1$. This suggests that the material reprocessing the outgoing light has significant angular momentum support, and that the particular line of sight through which \psone was observed contained only a fraction of the total mass accreted, $\lesssim 10^{-2} \msun$. For the power-law evolution of the reprocessing component (Equation \ref{eq:rph}), we find that $l = 2.76$, which indicates that the photosphere of the reprocessing component evolves significantly during the encounter, $m = -1.18$ (Equation \ref{eq:powlaws}). The fact that the highest-likelihood models deviate from $m = 0$ indicates that the distribution of matter in radius and height may be non-trivial, or that the ionization state may be changing as a function of time. The photosphere scale parameter $R_{\rm ph,0} = 0.17$ corresponds to a physical size of $9 \times 10^{14}$ cm at peak (approximately 50 times larger than pericenter distance $r_{\rm p} = 1.8 \times 10^{13}$ cm), the time evolution of which is shown in the right panel of Figure \ref{fig:threed}.

For the extinction in the host galaxy, we find that a column of $N_{\rm h} = 10^{21}$ cm$^{-2}$ is preferred, with a reddening parameter $R_{\rm v} = 6.5$. This value is somewhat higher than what is typically observed within the Milky Way ($R_{\rm v} = 3.1$), and is more representative of ``gray'' dust in which all wavelengths are absorbed equally. We verified that such a gray opacity is necessary by running a separate MCMC in which we fitted the extinction in each band independently, finding that the extinction in Pg is only 0.17 magnitudes greater Pz. Such values of $R_{\rm v}$ have been observed outside of the Milky Way \citep[see e.g.][]{Falco:1999ip}, and are typical of dense molecular clouds \citep{Draine:2003di}. Another possibility that our generalized model simply does not produce enough UV photons, necessitating a gray opacity law to compensate.

\begin{figure}
\centering\includegraphics[width=\linewidth,clip=true]{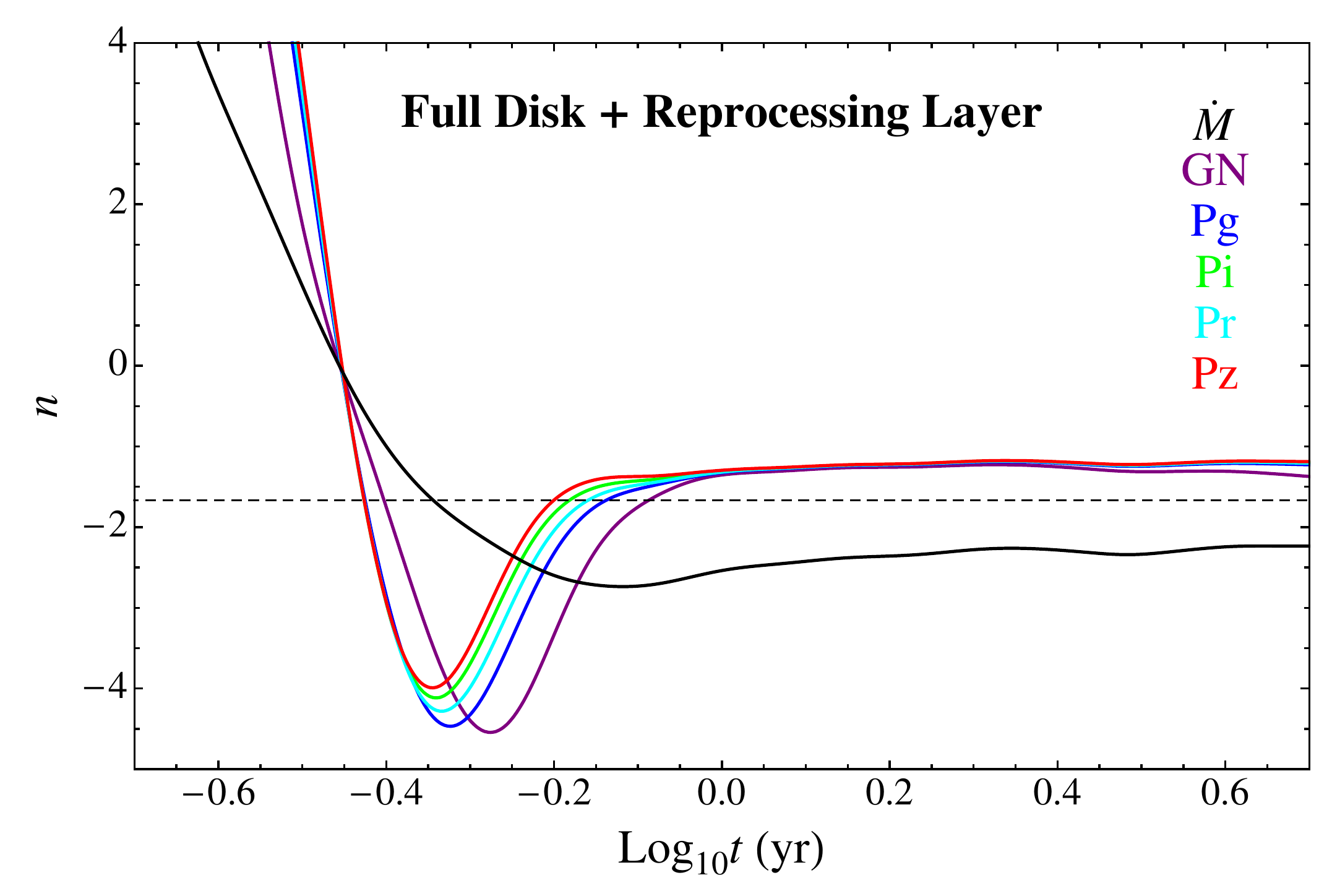}
\caption{Power-law index $n \equiv \partial \ln X / \partial \ln t$ for the highest-likelihood model shown in the right-hand panel of Figures \ref{fig:fits} and \ref{fig:seds} (corresponding to the generalized model), where $X$ is a placeholder for either the fallback rate $\md$ or the amount of flux incident on the detector in a given filter. The solid black curve shows the fallback rate $\md$, which is assumed in our model to be proportional to the bolometric luminosity, whereas the dashed black curve shows $n = -5/3$, the power-law index expected for the canonical constant-density star. The colored curves show $n$ for each band \psone was observed. Note that there is little color evolution at early times through peak (at $n = 0$), but some color evolution at late times.}
\label{fig:bandsn}
\end{figure}

Lastly, we find that the model requires $\sigma_{\rm v} = 0.05$ magnitudes of intrinsic variability, about double that expected for a steadily-accreting black hole with mass $10^{7} M_{\odot}$ \citep{Kelly:2011ku},
\begin{align}
\sigma_{\rm v} &= t_{\rm H} \zeta^{2}\nonumber\\
&= 0.0253^{+0.071}_{-0.038} M_{7}^{-0.19 \pm 0.78}\label{eq:sigmav},
\end{align}
where $t_{\rm H}$ is the timescale of the break in the PSD, and $\zeta$ is the square root of the variability amplitude measured at the break. This is surprisingly small given the potentially chaotic nature of the accretion process, and suggests that the accretion process is smooth and regular, with no major changes in global structure over short timescales.

In Figure \ref{fig:bandsn} we show the power-law index $n$ for the feeding rate $\md$ and luminosity measured in each of the {\it Pan-STARRS} and {\it GALEX} bands. It is clear that $\md$ does not asymptotically approach $n = -5/3$ for the most-likely model, as is expected given that the asymptotic $n$ ranges from $-1.4$ to $-2.2$ for MS disruptions \citep{Guillochon:2013jj}. The individual bands also deviate from $-5/3$ in the asymptotic limit, which again is not surprising given that the flux is a given band depends on the photosphere temperature \citep{Strubbe:2009ek}, which in our models varies as a function of time.

We find that our highest-likelihood models with and without the BLR constraints are very similar to one another. In Figure \ref{fig:fit-dists}, we present the posteriors of four fundamental parameters ($M_{\rm h}$, $M_{\ast}$, $a_{\rm spin}$, and $\beta$), both with (red) and without (blue) the BLR constraints, and find that difference in the posteriors is on the same order as the scatter about the median. This suggests that the timescale, luminosity, and color of \psone are sufficient to constrain most of the physical parameters of an event, whereas the BLR constraints can be used as a sanity check to ensure there is no discrepancy between the existent BLR emission regions and the observed spectra.

\section{Discussion}\label{sec:discussion}

\subsection{Arguments against the helium star interpretation}

The discovery of a flare with no noticeable hydrogen features certainly hints at the possibility that the disrupted star may have been relatively devoid of hydrogen. Aside from the hypothesis presented in the previous sections, there are other reasons to believe why the helium-rich progenitor scenario might be unlikely.

Firstly, helium-rich stars are rare in the universe. The known candidates are SdB/SdO stars \citep[$\Sim 10^{6}$ in the MW,][]{Han:2003hl}, helium WDs \citep[$\Sim 10^{7}$ in the MW, ][]{Nelemans:2001ia}, and WR stars \citep[$\Sim 10^{4}$ in the MW,][]{vanderHucht:2001gf}. While there is some evidence that the mass function around SMBHs is not well-represented by a canonical IMF \citep{Bartko:2010hh}, it seems unlikely that the numbers of these stars could be increased by the factor of $\Sim 10^{4}$--$10^{7}$ required to plausibly explain why the first well-resolved TDE happened to be a helium-rich star.

A second possibility is that the helium-rich star comes as the result of the previous interaction of a giant star with the SMBH, or potentially through a collision between the giant and a more compact stellar object \citep{Davies:1991hp}. However, in both of these cases, hydrogen is not completely removed from the star. In fact, even for deep tidal encounters, the core tends to retain an atmospheric mass of hydrogen comparable to its own mass \citep{MacLeod:2012cd}. Additionally, giant stars do not frequently get deposited into highly-bound orbits in which the core itself is likely to be disrupted, as the densities of their cores are $\gtrsim 10^{3}$ times larger than their envelopes, and their orbital migration into the loss-cone is largely dictated by diffusion \citep{Wang:2004jy,MacLeod:2012cd,MacLeod:2013dh}.

We can thus conclude that while helium-rich disruptions will occur occasionally, they will not be the dominant contributor to the rate, and as a result, it is highly unlikely by chance that these disruptions would be among the first to be observed.

\begin{figure}
\centering\includegraphics[width=\linewidth,clip=true]{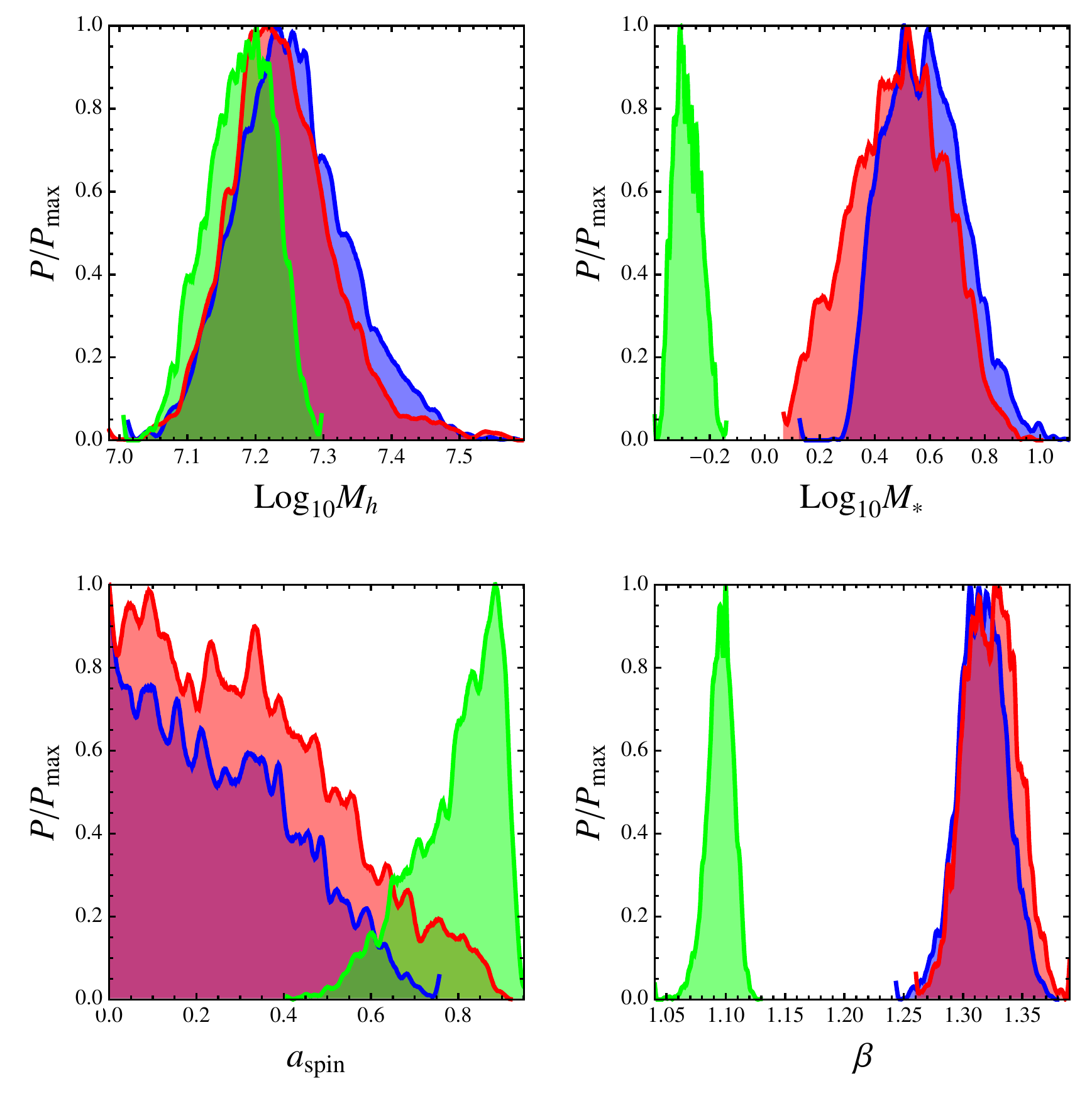}
\caption{Posterior distributions of $M_{\rm h}$, $M_{\ast}$, $a_{\rm spin}$, and $\beta$ for \psone. Within each panel are the probability $P$ scaled to the maximum probability $P_{\max}$, with the red curves showing the posteriors when no BLR constraints and no priors are included, the blue curves showing the posteriors when BLR constraints alone are included, and the green curves showing the posteriors when BLR constraints are combined with priors for $M_{\ast}$ and $\beta$. The BLR constraints alone (red) are only slightly different from posteriors calculated with no constraints (blue), but the inclusion of a prior on $M_{\ast}$ and $\beta$ suggests that the star that was disrupted was lower in mass ($0.5 M_{\odot}$ vs. $4 M_{\odot}$) and that the black hole was rapidly spinning (0.9 vs. 0.3).}
\label{fig:fit-dists}
\end{figure}

\subsection{Inclusion of priors}

In the previous sections, we did not make any assumptions about the distribution of any of our input parameters, but in reality these parameters are likely to have non-flat priors. For example, the distribution of stars around SMBHs is likely to possess a current mass function (CMF) that is strongly related to the initial mass function (IMF), which would suggest that the most likely stars to be disrupted are those with $M_{\ast} \sim 0.1 M_{\odot}$ \citep{Kroupa:2001ki}, in contrast to our unconstrained fits in which $M_{\ast} \simeq 4 M_{\odot}$. Additionally, we might expect that that grazing encounters (e.g. small $\beta$) should outnumber deep encounters \citep{Frank:1976tg}, and the black hole mass should follow established $M_{\rm h}$-$L$ relations \citep{Graham:2013fc}. The green posteriors in Figure \ref{fig:fit-dists} show how our most-likely parameters change when priors on $M_{\ast}$ \citep{Kroupa:2001ki} and $\beta$ (${\rm Prob.} \propto \beta^{-2}$) are included in our MLA, yielding a lower mass star with $M_{\ast} \simeq 0.5 M_{\odot}$ that was disrupted by a rapidly-spinning black hole ($a_{\rm spin} \simeq 0.9$). We find that the quality of the fit is only slightly poorer when priors are included, suggesting that a low-mass stellar disruption can fit equally well within the context of our generalized model.

However, each of these priors has a great deal of uncertainty associated with it. In our own galactic center, it is not clear if the distribution of stars is similar to the general IMF observed in the field, especially given the prevalence of short-lived B-stars within several lt-days of the black hole \citep{Gillessen:2009fn}. These stars may have been deposited through binary disruption \citep{Ginsburg:2006jg}, or through a disk \citep{Madigan:2011ej}, both of which would lead to different distribution in $\beta$ than what would be produced by a steady-state, spherically symmetric cluster around the black hole \citep[\`{a} la][]{Wang:2004jy}. Lastly, while a clear trend has been demonstrated between the luminosity of the host galaxy and the mass of its black hole, there is significant scatter about this trend \citep{Gultekin:2009hj}. In principle, once a significant number of TDEs have been identified, these distributions can be determined from the collection of fits to all disruptions, which could potentially improve the accuracy of parameter estimations of future events.

\subsection{How BLRs can help us understand TDEs}

The (non-)existence of various lines in spectra acquired of TDEs can be used to great effect to constrain the parameter space of allowed encounters for any particular event. These features enable one to place a time-dependent size constraint on the size of the debris structure resulting from a tidal disruption, which is directly related to the combination of three parameters: $M_{\rm h}$, $M_{\ast}$, and $\beta$ (Equation \ref{eq:ro}). In this paper, we have focused specifically on \heii and \ha in regards to \psone, but our technique could be used in general with other emission lines. \psone appears to originate from a moderately-massive SMBH, but disruptions of stars by more or less massive black holes would respectively produce larger or smaller structures from which emission lines could be produced. As an example, the disruption of the same star by a $10^{8} M_{\odot}$ may show \hb and \hei features early, with \ha appearing later, whereas a disruption by a $10^{6} M_{\odot}$ black hole may never show any helium or hydrogen emission lines.

In the optical at $z = 0$, the number of strong emission lines is limited, but many more emission lines are available in the UV and X-ray where metals with larger ionization potentials begin to lose their electrons. These lines, which would be produced nearer to the SMBH, could potentially be used to constrain the size scale at early times, and would provide a spatial map of the accretion disk as it grows. It is critical that TDEs are identified early and followed up spectroscopically to obtain this valuable information.

\subsection{How TDEs can help us understand BLRs}

The BLR has long been used to measure the masses of black holes from the lag times observed in the response of various emission lines, which are thought to lie at various distances. However, there remains much uncertainty in these models, namely the form of the BLR itself. If the dissipation mechanism within the debris disks resulting from tidal disruptions is similar to the dissipation mechanism that controls angular momentum transport in steadily-accreting AGN, it is reasonable to expect that the two structures should have many similarities in terms of their density and temperature profiles, velocity structures, and in the components of the structure that conspire to produce the emergent light.

In this paper, we have made direct comparisons to BLRs in order to understand the emission features that are observed in a particular event. As we have shown, the dependence between the distance at which a particular emission line is produced and the flux originating from the central engine is even more exaggerated than in the case of steadily-accreting AGN, as some line-emitting regions do not exist at all due to the absence of mass beyond a certain distance. With a larger catalogue of TDEs, we can reverse the arguments presented here to learn more about the structure of the BLR present in TDE debris disks.

\subsection{Caveats and future directions}

At the time of this writing, \psone is the only event that is claimed to be a TDE and also captures the rise, peak, and decay of the flare. By capturing all three phases, and with the addition of spectroscopic information, this event provides significantly more information on the underlying mechanisms than the small number of poorly sampled UV/optical TDEs that only capture the decay phase and may have no spectroscopic data \citep{Gezari:2006gd,Gezari:2008iv,Cappelluti:2009jl,vanVelzen:2011gz,Cenko:2012fg}. While the models presented here provide compelling evidence of the similarities between steadily-accreting AGN and luminous flares resulting from the tidal disruptions of stars, there are many aspects that can be improved upon. Some uncertainties in the generalized model presented here, such as details on the viscous processes that govern accretion and how matter light is reprocessed, could potentially be resolved with a more-complete collection of well-sampled TDEs.

It is clear from our purely hydrodynamical simulations that mere gas dynamics is incapable of generating the necessary dissipation for high mass-ratio encounters, as we described in Section \ref{subsec:disenough}. This suggests that magnetohydrodynamical simulations that focus on the nozzle region need to be performed to examine the growth of the MRI, which by our estimate may be capable of providing the required dissipation. If this mechanism is incapable of operating, then it is possible that only deeply-penetrating encounters in which $r_{\rm p} \sim r_{\rm g}$ will yield rapidly-rising light curves.

A second critical uncertainty is our treatment of the reprocessing layer, which is inextricably linked to the BLR of TDE debris disks. In this work, we have presumed that this reprocessing layer is spherical, parameterized the amount of light absorbed by an average gray opacity, and have ignored potentially complex radiation transport and line-of-sight effects. It is also unlikely that the BLR relations we compare to here are identical for debris disks resulting from disruption, given their elliptical geometry and different radial mass distributions. While the scaling relations determined for steadily-accreting AGN are likely to be similar to TDE scaling relations, meaningful constraints on individual events can only be obtained by revising these relations to account for the differences.

Given a more-accurate prescription of how the viscous and reprocessing mechanisms operate, \tdefit can easily be improved to include these additional aspects of the problem, which can potentially yield accurate estimates of the parameters associated with individual disruption events. With a large library of TDEs, which will likely exist in the LSST era when potentially thousands of TDEs may be detected \citep{vanVelzen:2011gz}, it should be possible to obtain detailed demographics of the stellar clusters that surround SMBHs.

\subsection{Lessons Learned}

For the readers convenience, we summarize the main findings of this paper below.

\begin{enumerate}
\item
The unbound material, while ejected at high velocity from pericenter after a disruption, is gravitationally confined in the two directions transverse to its motion. This constricts the debris to a thin stream that presents a negligible surface area as compared to the emitting surface generated by the return of the stream to pericenter, and is unlikely to affect the flare's appearance.
\item
When material returns to pericenter, it is heated via hydrodynamical shocks, but this dissipation is likely insufficient to explain the tight relationship between $L$ and \smash{$\md$} for large-$q$ encounters. Additional dissipation via an MRI-like mechanism or through general relativistic precession is probably required to explain this observed relationship.
\item
A disk that is truncated at $2 r_{\rm t}$ fails drastically in explaining the observed flare, and cannot match the observed shape of the light curves without extreme color evolution.
\item
The light curve of \psone is well modeled by a single blackbody whose temperature evolves weakly in time, and whose size is tens of times larger than $r_{\rm t}$. We speculate that this distance is roughly co-spatial with the distance at which helium is doubly-ionized.
\item
The fact that \Heii emission lines are observed, but \ha and \hb are not, is consistent with the size constraint on the bound debris that is ejected from the nozzle point upon returning to pericenter. In general, the presence or absence of various emission lines can be used as a probe of the size of the elliptical debris disk.
\item
When prior information is not included, the parameters for \psone of our highest-likelihood fits indicate that a $4 M_{\odot}$ main-sequence star was disrupted by a $2 \times 10^{7} M_{\odot}$ black hole. We find that the inclusion of a reasonable prior on $M_{\ast}$ and $\beta$ yields a lower stellar mass, $0.5 M_{\odot}$. While there is uncertainty in the proper prior to use in a galactic center environment, both the fits with and without priors involve the disruption of a common star by a common SMBH with an impact parameter near the expected average \citep{DeColle:2012bq}, and thus TDEs of the kind we associate with \psone are likely to be among the most common sub-Eddington disruption events. However, given that we are analyzing a single event in this paper, we cannot eliminate the possibility that an event of this type was among the first observed due to observational bias. Once more well-sampled TDEs are available, a joint analysis of many events similar to what we perform here is required for a complete understanding of the demographics of tidal disruption.
\end{enumerate}

\acknowledgments We have benefited from many useful discussions with J. Arnold, J. Braithwaite, B. Cenko, R. da Silva, K. Denney, C. Dorman, R. Foley, D. Foreman-Mackey, S. Gezari, J. Goodman, J. Halpern, D. Kasen, S. Kulkarni, D. N. C. Lin, M. MacLeod, C. Matzner, C. Miller, M. Pessah, M. Rees, S. Rosswog, A. Socrates, L. Strubbe, J. Trump, and E. Zweibel. We thank the anonymous referee for their constructive comments and suggestions, and C. M. Gaskell for detailed comments. The software used in this work was in part developed by the DOE-supported ASCI/Alliance Center for Astrophysical Thermonuclear Flashes at the University of Chicago. Computations were performed on the UCSC Pleiades, Hyades, and Laozi computer clusters, and the NASA Pleiades computer cluster. We acknowledge support from the David and Lucille Packard Foundation, NSF grants PHY-0503584 and ST-0847563, and the NASA Earth and Space Science Fellowship (JG).

\bibliographystyle{apj}
\bibliography{/Users/james/Dropbox/library}

\end{document}